\documentclass[12pt]{article}
\usepackage{latexsym,epsfig,amssymb} 
\usepackage[usenames]{color}
\pdfoutput=1

\definecolor{Blue}{rgb}{0,0,1}
\definecolor{Red}{rgb}{1,0,0}

\newcommand{\ft}[2]{{\textstyle\frac{#1}{#2}}}

\def\bfone{\relax{\rm 1\kern-.35em 1}}

\newcommand{\be}{\begin{equation}}
\newcommand{\ee}{\end{equation}}
\newcommand{\ben}{\begin{displaymath}}
\newcommand{\een}{\end{displaymath}}
\newcommand{\bea}{\begin{eqnarray}}
\newcommand{\eea}{\end{eqnarray}}

\newcommand{\non}{\nonumber\\}
\newcommand{\bean}{\begin{eqnarray*}}
\newcommand{\eean}{\end{eqnarray*}}

\makeatletter
\@addtoreset{equation}{section}
\makeatother

\newdimen\squaresize \squaresize=12pt
\newdimen\thickness \thickness=0.7pt

\def\square#1{\hbox{\vrule width \thickness
   \vbox to \squaresize{\hrule height \thickness\vss
      \hbox to \squaresize{\hss#1\hss}
   \vss\hrule height\thickness}
\unskip\vrule width \thickness} \kern-\thickness}

\def\cut#1{\hbox{\vrule width-1 \thickness
   \vbox to \squaresize{\hrule height-1 \thickness\vss
      \hbox to \squaresize{\hss#1\hss}
   \vss\hrule height-1\thickness}
\unskip\vrule width +4 \thickness} \kern-\thickness}

\def\vsquare#1{\vbox{\square{$#1$}}\kern-\thickness}

\def\young#1{
\vbox{\smallskip\offinterlineskip \halign{&\vsquare{##}\cr #1}}}

\newcommand{\tinyyoung}[1]{
\squaresize=7pt \thickness=0.4pt \mbox{\tiny\young{#1}}
\squaresize=12pt \thickness=0.7pt}

\begin{document}

\thispagestyle{empty}

\begin{flushright}\small 
ENSL-00315624
\end{flushright}
%%%%%%%%%%%%%%%%%%%%%%%%%%%%
%%%%%%%%%%%%%%%%%%%%%%%%%%%%
%

\bigskip
\bigskip

\vskip 10mm
\begin{center}
  {\LARGE {\bf Lectures on Gauged Supergravity\\[1ex]
  and Flux Compactifications}}

\bigskip
\medskip

{\em given at the RTN Winter School on\\ Strings, Supergravity 
and Gauge Theories, CERN, January 2008.}

\end{center}

\line(1,0){420}

\vskip 4mm

\begin{center}
{{\bf Henning Samtleben}\\[1ex]
Universit\'e de Lyon, Laboratoire de Physique, ENS Lyon,\\ 
46 all\'ee d'Italie, F-69364 Lyon CEDEX 07, France \\
{\tt henning.samtleben@ens-lyon.fr}}
\vskip 4mm
\end{center}

\vskip2cm

\begin{center} {\bf Abstract } \end{center}
\begin{quotation}\noindent
The low-energy effective theories describing string compactifications 
in the presence of fluxes are so-called gauged supergravities: 
deformations of the standard abelian supergravity theories. 
The deformation parameters can be identified with the various 
possible (geometric and non-geometric) flux components. 
In these lecture notes we review the construction of gauged 
supergravities in a manifestly duality covariant way and 
illustrate the construction in several examples.
\end{quotation}

\newpage

\tableofcontents

\bigskip
\bigskip
\bigskip
\bigskip
\bigskip

%%%%%%%%%%%%%%%%%%%%%%%%%%%%%%%%%%%%%%%
%%%%%%%%%%%%%%%%%%%%%%%%%%%%%%%%%%%%%%%

\section{Introduction}
\label{sec:intro}

%%%%%%%%%%%%%%%%%%%%%%%%%%%%%%%%%%%%%%%
%%%%%%%%%%%%%%%%%%%%%%%%%%%%%%%%%%%%%%%

Gauged supergravities have first been constructed in the early 1980's 
upon reconciling four-dimensional supergravity with maximal number of supercharges
with the non-abelian gauge structure of Yang-Mills theories~\cite{deWit:1982ig}.
Soon after, the first construction was generalized to other (non-compact) gauge 
groups~\cite{Hull:1984vg}
and to higher dimensions~\cite{Gunaydin:1985cu},~\cite{Pernici:1984xx}.
To date, gaugings are the only known supersymmetric 
deformations of maximal supergravity 
with the non-abelian gauge coupling constant acting as
a deformation parameter.
In recent years, these theories have reappeared
in particular in the context
of flux compactifications, see~\cite{Grana:2005jc,Blumenhagen:2006ci} for reviews. 
Non-vanishing background fluxes for the higher-dimensional $p$-form
tensor fields and so-called geometric fluxes twisting the internal geometry
of the compactification manifold may likewise act as deformation parameters
in the effective four-dimensional field theory.
The resulting actions can be described in the framework of
gauged supergravities with the resulting gauge groups typically being
of the non-semisimple type.

In these lectures we will 
review the construction of gauged supergravities, i.e.\
we will address the problem of describing the general
deformation of supergravity theories by coupling the abelian vector fields
to charges assigned to the elementary fields.
The general picture is sketched in figure~\ref{fig:reduction}:
starting from eleven-dimensional supergravity~\cite{Cremmer:1978km} 
(or alternatively the ten-dimensional type IIB theory~\cite{Schwarz:1983wa,Howe:1983sra})
the maximal supergravities in all lower dimensions 
are obtained by dimensional reduction on torus manifolds $T^n$ (the vertical arrow).
Their characteristic properties include exceptionally large global symmetry groups
and abelian gauge groups; e.g.~for the maximal four-dimensional theory these
are a global ${\rm E}_7$ and a local ${\rm U}(1)^{28}$ symmetry, 
respectively~\cite{Cremmer:1979up}.
None of the matter fields are charged under the abelian gauge group, hence the name
of {\em ungauged} supergravity.
Another distinct feature of these theories is their maximally supersymmetric
Minkowski ground state in which all fields are massless.

\begin{figure}[bt]
\begin{center}
\resizebox{120mm}{!}{\includegraphics{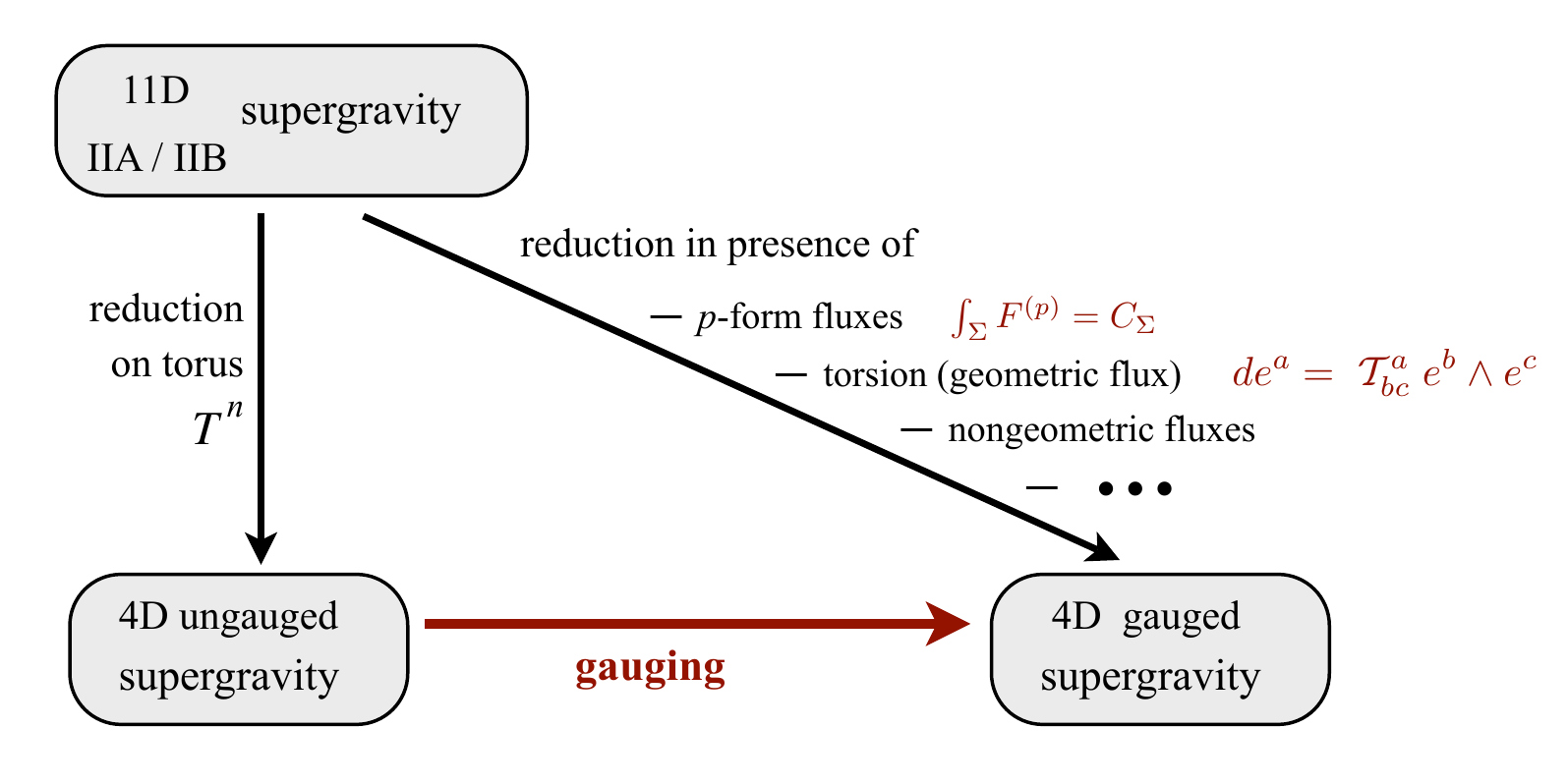}}
\caption{{\small Gauged supergravities and flux compactifications.}}
\label{fig:reduction}
\end{center}
\end{figure}

Instead, one may consider more complicated compactifications
(the diagonal arrow in figure~\ref{fig:reduction}),
in which e.g.\ the torus is replaced by manifolds with more structure
(such as spheres~$S^n$),
in which higher-dimensional $p$-form fields may acquire non-trivial
background fluxes,
in which the torus may be supplied with torsion, etc.
All these compactifications lead to more complicated
effective theories in four dimensions which typically come with
non-abelian gauge symmetries under which the matter fields are charged,
and which are referred to as {\em gauged} supergravities.
In contrast to their ungauged counterparts, these theories typically
come with a scalar potential which is a result of the more complicated
internal geometry. This is one of the reasons that 
has triggered the interest in these compactifications:
the scalar potential may support an effective cosmological constant, 
provide mass terms for the fields of the theory (moduli stabilization), 
describe scenarios of spontaneous supersymmetry breaking, etc.,
thereby accommodating many phenomenologically desirable properties.
Except for very few examples, these gauged supergravities do no longer
admit maximally supersymmetric groundstates in accordance with the
fact that the presence of non-vanishing background fluxes typically breaks
supersymmetry.

The most systematic approach to the construction and study of these gauged
supergravities is by considering them as deformations of the ungauged
theories obtained by simple torus reduction.
This is depicted by the horizontal arrow in figure~\ref{fig:reduction},
with the flux and geometric parameters acting as deformation parameters.
On the level of the four-dimensional theory, this construction
selects a subgroup~${\rm G}_0$ 
of the global symmetry group~${\rm G}$ of the ungauged theory
and promotes it to a local gauge symmetry 
by coupling it to the (formerly abelian) vector fields of the theory.
As a result, the matter fields of the theory are charged under the new
gauge symmetry.
The first example of such a construction was the ${\rm SO}(8)$ gauged theory 
of~\cite{deWit:1982ig} which describes the~$S^7$ compactification of 
eleven-dimensional supergravity, with ${\rm SO}(8)$ properly embedded 
into the global ${\rm E}_7$ symmetry of the ungauged theory.
In the context of flux compactifications, many other typically non-semisimple
gaugings of this theory have been identified, some of which we will describe in the last section.
Fortunately, all different gaugings can be described in a single covariant
construction that is based on the underlying global symmetry group~${\rm G}$ 
of the 
ungauged theory. This framework, first developed in the context of
three-dimensional supergravity~\cite{Nicolai:2000sc,Nicolai:2001sv}
and further shaped in~\cite{deWit:2002vt,deWit:2004nw,deWit:2005hv},
encodes the possible gaugings in an {\em embedding tensor}
that describes the embedding of the gauge group into the
global symmetry group, can be characterized group-theoretically,
and turns out to entirely parametrize the action of the gauged supergravity.
From the point of view of flux compactifications, this can be seen as a very
compact way to group all the different possible flux (or deformation) parameters
into a single tensorial object on which furthermore 
the action of the duality group is manifest. 
This will be the central theme of these lectures.

So far we have presented the picture for the supergravities
with maximal number of supercharges
and for definiteness we will throughout stick with the maximal
or half-maximal theories whose structures are very rigid due to 
the large underlying global symmetries. We should stress however, that
a large part of the structures and techniques to be presented directly apply to the
supergravities with lower number of supercharges.
E.g.\ in many applications the torus manifold in figure~\ref{fig:reduction}
would be replaced by a Calabi-Yau manifold such that the ungauged
four-dimensional supergravity is no longer maximal but has only ${\cal N}=2$ supersymmetry.
In complete analogy to the construction presented in the following, the effect
of non-vanishing background fluxes can be accommodated by gauging
certain global symmetries in these models leading to the same type of
gauged supergravity in four dimensions. This has been confirmed in
many explicit examples, see e.g.~\cite{Louis:2002ny,Grana:2006hr,DAuria:2007ay}.

As a last point we mention that gauged supergravities
have recently (re)appeared in other contexts as well
which we will not further discuss in these lectures.
Two important ones are the following:

\begin{itemize}

\item
The supergravity regime 
of the bulk theory in the AdS/CFT correspondence~\cite{Aharony:1999ti}
is generically described by a gauged supergravity.
It is due to its scalar potential that the theory can support
an AdS ground state. The corresponding gauge group is usually
compact and corresponds to the R-symmetry group of the boundary theory.
The prime-example is the five dimensional maximal ${\rm SO}(6)$ gauged
supergravity of~\cite{Gunaydin:1985cu} which describes IIB supergravity
compactified on AdS$_5\times S^5$. Its scalar potential encodes non-trivial 
information about the four-dimensional SYM boundary theory,
such as holographic RG flows and the anomalous conformal 
dimensions of operators~\cite{Girardello:1998pd,Freedman:1999gp,Bianchi:2001kw}.

\item
The structure of the gauged supergravities fits naturally with 
and gives further support to the proposals for the higher rank 
Kac-Moody symmetries
${\rm E}_{10}$~\cite{Damour:2002cu}
and ${\rm E}_{11}$~\cite{West:2001as},
conjectured to underlie supergravity and string theory,
As we shall discuss in section~\ref{sec:gauging}, 
the field content of gauged supergravities is typically larger as compared to the
ungauged theories, since the former naturally include a number of dual 
tensor fields and in particular the non-propagating 
antisymmetric $(D-1)$- and $D$-form tensor fields.
This larger field content is precisely in accordance with 
certain decompositions of the representations
of the extended infinite-dimensional 
Kac-Moody algebras~\cite{Riccioni:2007au,Bergshoeff:2007qi}.

\end{itemize}

The structure of these lectures is straightforward:
in section~\ref{sec:ungauged} we briefly review the structure of 
ungauged supergravities, in particular the role and the realization of 
their global symmetry groups ${\rm G}$.
In section~\ref{sec:gauging} we describe the gauging of these theories
in a formalism covariant under the symmetry group ${\rm G}$.
Finally, in section~\ref{sec:flux}, we discuss the higher-dimensional
origin of the gauged theories, in particular their application to 
the description of flux compactifications,
and illustrate the connection with several examples.

%%%%%%%%%%%%%%%%%%%%%%%%%%%%%%%%%%%%%%%
%%%%%%%%%%%%%%%%%%%%%%%%%%%%%%%%%%%%%%%

\section{Ungauged Supergravity ---  \mbox{Symmetries and Dualities}}
\label{sec:ungauged}

%%%%%%%%%%%%%%%%%%%%%%%%%%%%%%%%%%%%%%%
%%%%%%%%%%%%%%%%%%%%%%%%%%%%%%%%%%%%%%%

In this section, we collect some of the pertinent facts about
ungauged supergravity theories. 
The discussion will be rather brief and is not meant
to be an exhaustive introduction to these theories ---
for which we refer to the many excellent reviews in the literature,
see e.g.~\cite{VanNieuwenhuizen:1981ae,Tanii:1998px,Fre:2001jd,deWit:2002vz,VanProeyen:2003zj}.
Rather, we will here focus onto those elements that 
prove to be important for the subsequent construction of 
gauged supergravities,
notably the underlying structure of symmetries and dualities.
We will mainly restrict to the bosonic sector of these theories,
although its structure is of course to a large extent determined by 
the underlying supersymmetric extension.

The bosonic field content of standard supergravity theories consists of
the metric~$g_{\mu\nu}$, a set of scalar fields $\phi^{i}$, 
as well as vector fields $A_{\mu}^{M}$, 
and higher-rank antisymmetric $p$-forms $B_{\nu_1\dots\nu_p}^{I}\,$
of various ranks.
Their dynamics is described in terms of a Lagrangian
of the type
\bea
e^{-1}{\cal L}_{\rm bos} &=&
-\ft12R-\ft12\,G_{ij}(\phi)
\,\partial_{\mu}\phi^{i}\,\partial^{\mu}\phi^{j}
-\ft14\,{\cal M}_{MN}(\phi)\,F_{\mu\nu}^{M}\,F^{\mu\nu\,{N}}
-~ \cdots
\;,
\label{Lbos}
\eea
with $e=\sqrt{|{\rm det}\,g_{\mu\nu}|}$, and the abelian field strengths 
$F_{\mu\nu}^{\cal M}\equiv \partial^{\vphantom{\cal M}}_{\mu}A_{\nu}^{M}-\partial^{\vphantom{\cal M}}_{\nu}A_{\mu}^{M}$\,.
The dots here refer to kinetic terms for the higher-rank 
$p$-forms and to possible topological terms.
We will in these lectures always work in the Einstein frame, i.e.\
absorb possible dilaton prefactors of the Ricci scalar~$R$ by conformal
rescaling of the metric.

The form of the Lagrangian~(\ref{Lbos}) is essentially fixed by diffeomorphism
and gauge covariance (upon restricting the dynamics to two-derivative terms). 
The ``data'' that remain to be specified are
the scalar and the vector kinetic matrices $G_{ij}(\phi)$ and ${\cal M}_{MN}(\phi)$, 
respectively, as well as their counterparts for the higher-rank $p$-forms.
In general, the form of these scalar-dependent matrices is
highly constrained by supersymmetry. In the following we will mainly 
consider maximal and half-maximal supergravities 
(i.e.\ theories with 32 and 16 real supercharges, respectively)
for which the possible
couplings are extremely restrictive and organized by the structure
of an underlying global symmetry group~${\rm G}$.
Most of the discussion straightforwardly extends to theories with 
a lower number of supercharges, in particular in those cases in which 
the underlying global symmetry group is still sufficiently large.

In the rest of this section, we will describe how the global 
symmetry group restricts the field content and determines the 
Lagrangian~(\ref{Lbos}) without further explicit reference to supersymmetry. 
The $p$-forms and the scalar fields
of the theory transform in linear and non-linear
representations of~${\rm G}$, respectively,
whereas the metric $g_{\mu\nu}$ is left invariant by the action of ${\rm G}$.

%%%%%%%%%%%%%%%%%%%%%%%%%%%%%%%%%%%%%%%
\begin{table}[tb]
\begin{center}
{\small
\begin{tabular}{r||c|c|c|c|}
$D$ &${\rm G}_{\rm max}$  &${\rm K}_{\rm max}$  &${\rm G}_{\rm half-max}$
&${\rm K}_{\rm half-max}$     \\
\hline
9   & ${\rm GL}(2)$ & ${\rm SO}(2)$ & ${\rm GL}(1)\times {\rm SO}(1,1\!+\!n)$
& ${\rm SO}(1\!+\!n)$ \\
8   & ${\rm SL}(2)\! \times \!{\rm SL}(3)$  & ${\rm SO}(2)\! \times \!{\rm SO}(3)$& 
${\rm GL}(1)\times {\rm SO}(2,2\!+\!n)$ 
& ${\rm SO}(2)\! \times \!{\rm SO}(2\!+\!n)$\\
7   & ${\rm SL}(5)$  & ${\rm SO}(5)$ & ${\rm GL}(1)\times {\rm SO}(3,3\!+\!n)$
& ${\rm SO}(3)\! \times \!{\rm SO}(3\!+\!n)$\\
6   &  ${\rm SO}(5,5)$ & ${\rm SO}(5)\! \times \!{\rm SO}(5)$
& ${\rm GL}(1)\times {\rm SO}(4,4\!+\!n)$
& ${\rm SO}(4)\! \times \!{\rm SO}(4\!+\!n)$\\
5   & E$_{6(6)}$  & ${\rm USp}(8)$ & ${\rm GL}(1)\times {\rm SO}(5,5\!+\!n)$
& ${\rm SO}(5)\! \times \!{\rm SO}(5\!+\!n)$\\
4   & E$_{7(7)}$  & ${\rm SU}(8)$ &  ${\rm SL}(2)\times {\rm SO}(6,6\!+\!n)$
& ${\rm SO}(2)\! \times \!{\rm SO}(6)\! \times \!{\rm SO}(6\!+\!n)$\\
3   & E$_{8(8)}$  & ${\rm SO}(16)$ & ${\rm SO}(8,8\!+\!n)$
& ${\rm SO}(8)\! \times \!{\rm SO}(8\!+\!n)$\\
2   & E$_{9(9)}$  & ${\rm K}({\rm E}_9)$& ${\rm SO}(8,8\!+\!n)^{(1)}$
& ${\rm K}({\rm SO}(8,8\!+\!n)^{(1)})$
\end{tabular}
}
\end{center}
\caption{\small
Global symmetry groups ${\rm G}$ and their compact subgroups ${\rm K}$
in maximal and half-maximal supergravity in various dimensions. 
The subscripts in parentheses ${\rm E}_{N(N)}$ specify the particular real
form of the exceptional groups; for maximal supergravity this is
always the split form, i.e.\ the maximally non-compact form of the group.
For $D=2$, the groups 
${\rm E}_{9(9)}$ and ${\rm SO}(8,8\!+\!n)^{(1)}$ refer to the
(centrally extended) affine extensions of the groups
${\rm E}_{8(8)}$ and ${\rm SO}(8,8\!+\!n)$, respectively,
${\rm K}({\rm G})$ denotes their maximal compact subgroup.
}\label{table:groups}
\end{table}
%%%%%%%%%%%%%%%%%%%%%%%%%%%%%%%%%%%%%%%

%%%%%%%%%%%%%%%%%%%%%%%%%%%%%%%%%%%%%%%
\subsection{Scalar sector}
\label{subsec:scalar}
%%%%%%%%%%%%%%%%%%%%%%%%%%%%%%%%%%%%%%%

The scalar fields $\phi^i$ in (half-)maximal supergravity 
are described by a ${\rm G}/{\rm K}$ coset space sigma-model,
where ${\rm G}$ is the global symmetry group of the theory, collected in 
table~\ref{table:groups} for various dimensions,
and ${\rm K}$ is its maximal compact subgroup.
A convenient formulation of this sigma-model has the scalar fields 
parametrize a ${\rm G}$-valued matrix ${\cal V}$ 
(evaluated in some fundamental representation of ${\rm G}$)
and makes use of the left-invariant current
\bea
J_{\mu}&=& {\cal V}^{-1}\,\partial_{\mu}{\cal V} 
~\in~{\mathfrak{g}}~\equiv~ {\rm Lie}\,{\rm G} \;.
\eea
In order to accommodate the coset space structure,
$J_{\mu}$ is decomposed according to
\bea
J_{\mu} &=& Q_{\mu}+P_{\mu}
\;,\qquad
Q_{\mu}\in\mathfrak{k}\;,\;\; 
P_{\mu}\in\mathfrak{p}\;,\;\;
\eea
where $\mathfrak{k}\equiv{\rm Lie}\,{\rm K}$ and $\mathfrak{p}$
denotes its complement, i.e.\ $\mathfrak{g}=\mathfrak{k}\perp \mathfrak{p}$,
orthogonal w.r.t.\ the Cartan-Killing form.
The scalar Lagrangian is given by
\bea
{\cal L}_{{\rm scalar}} &=& -\ft12\,e\,{\rm Tr}\,(P_{\mu} P^{\mu})
\;.
\label{LPP}
\eea
It is invariant under global ${\rm G}$ and local ${\rm K}$ transformations
acting as
\bea
\delta\,{\cal V} &=& \Lambda\,{\cal V} - {\cal V}\,k(x)
\;,\qquad
 \Lambda\in\mathfrak{g}\;, \;\; k(x)\in\mathfrak{k} \;,
\label{symmGK}
\eea
on the scalar matrix ${\cal V}$. Under these symmetries 
the currents $Q_{\mu}$ and $P_{\mu}$ transform according to
\bea
\delta\,Q_{\mu} &=& -\partial_{\mu}k +[\,k,Q_{\mu}]\;,\qquad
\delta\,P_{\mu} ~=~ [\,k,P_{\mu}]
\;,
\eea
showing that the composite connection $Q_{\mu}$ behaves as a gauge field
under ${\rm K}$. As such it plays the role of a connection in 
the covariant derivatives
of the fermion fields which transform 
linearly under the local ${\rm K}$ symmetry,
e.g.\ 
\bea
D^{\vphantom{i}}_{\mu}\psi_{\nu}^{i} &\equiv&
\partial^{\vphantom{i}}_{\mu}\psi_{\nu}^{i}
-\frac14\omega_{\mu}{}^{ab}\,\gamma_{ab}\,\psi_{\nu}^{i}
-(Q_{\mu})_{k}{}^{i}\,\psi_{\nu}^{k}
\;,
\eea
for the gravitinos $\psi^{i}_{\nu}$, etc.
Likewise, one defines $D_\mu{\cal V}\equiv
\partial_\mu{\cal V}-Q_\mu = P_\mu$ for the scalar matrix~${\cal V}$.
The current $P_\mu$ on the other hand transforms in a linear
representation of~${\rm K}$, builds the ${\rm K}$-invariant kinetic term~(\ref{LPP}) 
and may be used to construct
${\rm K}$-invariant fermionic interaction terms in the Lagrangian.

The two symmetries~(\ref{symmGK}) extend to the 
entire supergravity field content and play a crucial role in 
establishing the full supersymmetric action.
They will furthermore be of vital importance in
organizing the construction
of the gauged theories described in the next section.
The global $\mathfrak{g}$ transformations may be expanded as
$ \Lambda= \Lambda^\alpha t_\alpha$ into a basis of generators $t_\alpha$
satisfying standard Lie-algebra commutation relations
\bea
{}[t_\alpha,t_\beta] &=& f_{\alpha\beta}{}^\gamma\,t_\gamma \;,
\eea
with structure constants $f_{\alpha\beta}{}^{\gamma}$.

The local ${\rm K}$ symmetry is not a gauge symmetry 
associated with propagating gauge fields
(the role of the gauge field is played by the composite 
connection $Q_\mu$),
but simply takes care of the redundancy in parametrizing the coset 
space ${\rm G}/{\rm K}$.
It is indispensable for the description of the 
fermionic sector,
with the fermionic fields transforming in linear representations 
under ${\rm K}$. In particular, the scalar matrix ${\cal V}$
transforming as~(\ref{symmGK}) 
can be employed to describe couplings between 
bosonic and fermionic fields, transforming under ${\rm G}$ 
and ${\rm K}$, respectively. To make this more explicit, 
it is useful to express~(\ref{symmGK}) in indices as
\bea
\delta{\cal V}_M{}^{\underline N} &=& 
\Lambda^\alpha\,(t_{\alpha})_M{}^K\,
{\cal V}_K{}^{\underline N} - 
{\cal V}_M{}^{\underline K}\,k_{\underline{K}}{}^{\underline N}
\;,
\eea
with ${\rm G}$-generators $(t_{\alpha})_M{}^K$ and the underlined
indices ${\underline{K}}, {\underline{N}}$ referring to their 
transformation behavior under the subgroup ${\rm K}$.
The matrix ${\cal V}_M{}^{\underline N}$ allows to construct
couplings of e.g.\ a bosonic field strength $F^M_{\mu\nu}$
transforming in the associated fundamental representation 
of ${\rm G}$ to the fermionic fields according to (schematically)
\bea
F^{M}\,{\cal V}_M{}^{\underline{N}}\,(\bar\psi \,\psi)_{\underline{N}}
\;,
\qquad
{\rm etc.}
\;,
\eea
where $(\bar\psi \,\psi)_{\underline{N}}$ denotes the projection of the fermionic 
bilinear onto some ${\rm K}$-subrepresen\-tation in the corresponding 
tensor product of ${\rm K}$-representations.

It is often convenient to
fix the local ${\rm K}$ symmetry by adopting a particular 
form of the matrix ${\cal V}$,
i.e.\ choosing a particular set of coset representatives. 
In this case, any global ${\rm G}$-transformation in~(\ref{symmGK})
needs to be accompanied by a compensating ${\rm K}$-transformation
\bea
\delta\,{\cal V} &=&  \Lambda\,{\cal V} - {\cal V}\,k_\Lambda
\;,
\label{nonlinear}
\eea
where $k_\Lambda$ depends on $ \Lambda$ (and on ${\cal V}$) 
in order to restore the particular gauge choice,
i.e.\ to preserve the chosen set of coset representatives.
This defines a non-linear representation of ${\rm G}$ on the
$({\rm dim}\,{\rm {G}} - {\rm dim}\,{\rm {K}})$ coordinates of 
the coset space, i.e.\ on the physical scalar fields.
Likewise, it provides a non-linear realization of the group ${\rm G}$
on the fermion fields via the compensating transformation~$k_\Lambda$.
Two prominent gauge fixings are the following:
\begin{itemize}

\item
{\em unitary gauge:} in which the matrix ${\cal V}$ is taken of the form
\bea
{\cal V} &=& {\rm exp}\left\{ \phi^a \, Y_a  \right\}
\;,
\eea
where the non-compact generators $Y_{a}$ span the space $\mathfrak{p}$.
In this gauge, the $\phi^a$ transform in a linear representation 
of ${\rm K}\subset{\rm G}$, thus global ${\rm K}$-invariance of the 
Lagrangian remains manifest.
The current $P_{\mu}=P^a_{\mu}\,Y^a$ takes the form 
$P^a_\mu = \partial_\mu\phi^a + \dots$, 
where dots refer to higher order contributions. This shows that 
the kinetic term~(\ref{LPP}) is manifestly ghost-free with 
$G_{ab}(\phi)\propto\delta_{ab}+\dots$. It is here that
the importance of ${\rm K}$ being the maximal compact 
subgroup of ${\rm G}$ shows up.

\item
{\em triangular gauge:} in which the matrix ${\cal V}$ is taken of the form
\bea
{\cal V} &=& 
{\rm exp}\left\{ \phi^m \, N_m  \right\}\,{\rm exp}\left\{ \phi^\lambda \, h_\lambda  \right\}
\;,
\label{triangular}
\eea
where $\lambda=1, \dots, {\rm rank}\,{\rm G}$, labels a set of 
Cartan generators $h_\lambda$ of $\mathfrak{g}$ and 
the $N_m$ form a set of nilpotent generators 
such that the algebra spanned by $\{h_\lambda, N_m\}$ constitutes a Borel
subalgebra of $\mathfrak{g}$.
With suitable choice of the Borel subalgebra, it is in this gauge that 
a possible higher-dimensional origin of the theories 
becomes the most transparent. 
The grading associated with the chosen Borel subalgebra is 
related to the charges of the fields under rescaling of the volume 
of the internal compactification manifold.
We shall illustrate this in section~\ref{sec:flux} in several examples,
see~\cite{Cremmer:1997ct} for a systematic discussion for the maximal
theories and their eleven-dimensional origin.
\end{itemize}

E.g.\ the scalar sector of the maximal (${\cal N}=8$) supergravity in 
$D=4$ space-time dimensions
is described by the coset space ${\rm E}_{7(7)}/{\rm SU}(8)$.
The eleven-dimensional origin of the fields can be identified
in the triangular gauge associated with the ${\rm GL}(7)$ 
grading of ${\rm E}_{7(7)}$.
A type IIB origin of the fields on the other hand is identified
in the triangular gauge associated with a particular 
${\rm GL}(6)\times {\rm SL}(2)$ grading. 
We shall come back to this in section~\ref{sec:flux}.
For the half-maximal (${\cal N}=4$) supergravity in $D=4$ dimensions,
the scalar sector is described by the coset space
\bea
{\rm G}/{\rm K} &=&
{\rm SL}(2)/{\rm SO}(2) \times\;
{\rm SO}(6,6\!+\!n)\Big/({\rm SO}(6)\!\times\!{\rm SO}(6\!+\!n))
\;,
\label{half-maximal}
\eea
where $n$ refers to the number of vector multiplets in the ten-dimensional
type-I theory, from which this theory is obtained by torus reduction. The 
ten-dimensional origin of the fields is identified in the triangular gauge
associated with the ${\rm GL}(6)$ grading of ${\rm SO}(6,6)$.

In order to construct the full supersymmetric action
of the theory, it is most convenient to keep the local 
${\rm K}$ gauge freedom. When discussing only the bosonic sector
of the theory, it is often functional to formulate
the theory in terms of manifestly ${\rm K}$-invariant objects.
E.g.\ the scalar fields can equivalently be described in terms 
of the positive definite symmetric scalar matrix ${\cal M}$ defined by
\bea
{\cal M} &\equiv& {\cal V}\,  \Delta\,{\cal V}^{\rm T}
\;,
\label{defM}
\eea
where $\Delta$ is a constant ${\rm K}$-invariant positive definite matrix
(e.g.\ for the coset space ${\rm SL}(N)/{\rm SO}(N)$,
with ${\cal V}$ in the fundamental representation,
$\Delta$ is simply the identity matrix).
The matrix ${\cal M}$ is manifestly ${\rm K}$-invariant and
transforms under ${\rm G}$ as
\bea
\delta{\cal M} &=& \Lambda\,{\cal M} + {\cal M} \,\Lambda^{\rm T}
\;,
\label{transM}
\eea
while the Lagrangian~(\ref{LPP}) takes the form
\bea
{\cal L}_{{\rm scalar}} &=& \ft18\,{\rm Tr}\,
(\partial_{\mu}{\cal M}\:\partial^{\mu}\!{\cal M}^{-1})
\;.
\eea

To finish this section, let us evaluate the general formulas for
the simplest non-trivial coset-space ${\rm SL}(2)/{\rm SO(2)}$
which appears in the matter sector of several supergravity theories.
With the $\mathfrak{sl}(2)$ generators given by
\bea
{\bf h}=
\left(
\begin{array}{cc}
1&0\\
0&-1
\end{array}
\right)
\;,\qquad
{\bf e}=
\left(
\begin{array}{cc}
0&1\\
0&0
\end{array}
\right)
\;,\qquad
{\bf f}=
\left(
\begin{array}{cc}
0&0\\
1&0
\end{array}
\right)
\;,
\eea
the matrix ${\cal V}$ in triangular gauge~(\ref{triangular})
is given as
\bea
{\cal V} &=& 
{\rm e}^{C\,{\rm\bf e}}\,{\rm e}^{\phi \,{\rm\bf h}}
~=~
\left(
\begin{array}{cc}
1 & C \\
0 & 1
\end{array}
\right)
\left(
\begin{array}{cc}
{\rm e}^\phi & 0 \\
0 & {\rm e}^{-\phi}
\end{array}
\right)
\;.
\eea
Evaluating the non-linear realization~(\ref{nonlinear})
of ${\rm SL}(2)$ on these coset coordinates $C$, $\phi$ leads to
\bea
\delta_{\bf h}\,\phi=1
\;,\quad
\delta_{\bf h}\,C=2C
\;,\qquad
\delta_{\bf e}\,C=1
\;,\qquad
\delta_{\bf f}\,\phi=-C
\;,\quad
\delta_{\bf f}\,C={\rm e}^{4\phi}-C^2
\;.
\label{hef}
\eea
This shows that ${\bf h}$ acts as a scaling symmetry
on the fields, whereas ${\bf e}$ acts as a shift symmetry on $C$,
and ${\bf f}$ is realized non-linearly.
This toy example exhibits already all the generic features
of the global ${\rm G}$-symmetries 
that we will meet in more generality in section~\ref{subsec:origin} below.

The matrix ${\cal M}$ for this model can be computed from~(\ref{defM}) 
with $\Delta=\mathbb{I}_2$, and is most
compactly expressed in terms of 
a complex scalar field ${\tau}=C+i {\rm e}^{2\phi}$, giving rise to
\bea
{\cal M}&=&\frac1{\Im\tau}\,
\left(
\begin{array}{cc}
|\tau|^2 & \Re\tau\\
\Re\tau & 1
\end{array}
\right)\;,
\label{Mex}
\eea
while the kinetic term~(\ref{LPP}) takes the form
\bea
e^{-1}\,{\cal L}_{\rm scalar} &=&
-\partial_\mu\phi \,\partial^\mu\phi
- \ft14\,e^{-4\phi} \,\partial_\mu C \,\partial^\mu C
~=~
-\frac1{4(\Im\tau)^2}\,\partial_\mu\tau \,\partial^\mu\tau^* 
\;.
\eea
It is manifestly invariant under the scaling and shift symmetries 
of~(\ref{hef}), whereas invariance under the non-linear action of ${\bf f}$ is not
obvious and sometimes referred to as a hidden symmetry. 
In terms of $\tau$, the action of a finite ${\rm SL}(2)$ 
group transformation can be given in the 
compact form
\bea
\tau &\rightarrow& \frac{a\tau+b}{c\tau +d}\;,\qquad
\mbox{for}\quad
{\rm exp}(\Lambda) ~=~ 
\left(
\begin{array}{cc}
a & b \\
c & d
\end{array}
\right)~\in~{\rm SL}(2)
\;.
\eea

%%%%%%%%%%%%%%%%%%%%%%%%%%%%%%%%%%%%%%%
\subsection{Vectors and antisymmetric $p$-forms}
\label{subsec:vector}
%%%%%%%%%%%%%%%%%%%%%%%%%%%%%%%%%%%%%%%

The $p$-forms in ungauged supergravity transform in 
(typically irreducible) linear representations of the global symmetry group ${\rm G}$.
E.g.\ while the scalar fields transform under ${\rm G}$ as (\ref{symmGK}),
the transformation of the vector fields $A_\mu^M$ 
($M=1, \dots, n_{\rm v}$) is given by
\bea
\delta A_\mu^M &=& -\Lambda^\alpha\, (t_\alpha)_N{}^M\,A_\mu^N
\;,
\label{actionA}
\eea
where $(t_\alpha)_N{}^M$ denote the generators of $\mathfrak{g}$ in a fundamental representation
${\cal R}_{\rm v}$ with $\mbox{${\dim}\,{\cal R}_{\rm v}=n_{\rm v}$}$.
Similarly, the higher-rank $p$-forms transform in particular
representations of ${\rm G}$. The $p$-form field content of the 
ungauged maximal supergravities in various dimensions
is determined by supersymmetry and collected in table~\ref{table:pforms}.

\begin{table}[tb]
\centering
\begin{tabular}{r|c||c|c|c|c|}
$D$ &${\rm G}$& $0$& $1$&$2$&$3$  \\   \hline
9   & ${\rm GL}(2)$  &
${\bf 1}^0 + {\bf 3}^0 - 1$ 
& ${\bf 1}^{-4} + {\bf 2}^{+3}$ & ${\bf 2}^{-1}$ & ${\bf 1}^{+2}$\\
8   & ${\rm SL}(2)\! \times \!{\rm SL}(3)$  &
$({\bf 3}\!-\!1,{\bf 1})\!+\!({\bf 1},{\bf 8}\!-\!3)$ & 
 $({\bf 2},{\bf 3}')$&  $({\bf 1},{\bf 3})$&  $(\framebox{{\bf 2}},{\bf 1})$

 \\
7   & ${\rm SL}(5)$ &${\bf 25}-15$ & ${\bf 10}'$  & ${\bf 5}$ &   \\
6  & ${\rm SO}(5,5)$ &${\bf 45}-30$ & ${\bf 16}_c$ & $\framebox{\bf 10}$ &\\
5   & ${\rm E}_{6(6)}$ &${\bf 78}-36$ & ${\bf 27}'$ & & \\
4   & ${\rm E}_{7(7)}$ &${\bf 133}-63$ & $\framebox{\bf 56}$ & & \\
3   & ${\rm E}_{8(8)}$ & ${\bf 248}-120$ & & &
\\ 
\end{tabular}
\caption{\small
The $p$-form field content in ungauged maximal supergravity
organizes into ${\rm G}$-representations.
The physical scalars ($p=0$) descend from the 
adjoint representation of ${\rm G}$ upon eliminating 
$({\rm dim}\,{\rm K})$ of them by fixing the local ${\rm K}$ freedom,
cf.~(\ref{nonlinear}).
The framed representations appearing in the even dimensions
refer to the peculiarity concerning the 
$(D/2-1)$ forms, of which only half appear 
in the Lagrangian and carry propagating degrees of freedom, 
as discussed in subsection~\ref{subsec:even} below.
}\label{table:pforms}
\end{table}

An invariant action for the vector fields is given by
\bea
{\cal L}_{\rm kin} &=&
-\ft14\,e\,{\cal M}_{MN}\,F_{\mu\nu}^M\,F^{\mu\nu\,N}
\;,
\label{Lkin}
\eea
with the abelian field strength 
$F_{\mu\nu}^{\cal M}\equiv 
\partial^{\vphantom{\cal M}}_{\mu}A_{\nu}^{M}-
\partial^{\vphantom{\cal M}}_{\nu}A_{\mu}^{M}$
and the scalar dependent positive definite 
matrix ${\cal M}_{MN}$ defined in~(\ref{defM}).
This action is manifestly invariant under ${\rm G}$ 
with ${\cal M}$ transforming as~(\ref{transM}).
To be precise, the action~(\ref{Lkin}) is only relevant for
the vector fields in $D>4$ dimensions, while in $D=4$ 
space-time dimensions the story is somewhat more
complicated as a consequence of electric/magnetic duality
as we shall briefly review in the next subsection.
Similarly, the kinetic terms for higher-rank $p$-forms $B^I_{\nu_1\dots\nu_{p}}$
(with $p<[(D\!-\!1)/2]$) are governed by the 
positive definite scalar matrices~(\ref{defM})
evaluated in the corresponding representations
\bea
{\cal L}_{\rm kin} &=&
-\ft1{2(p+1)!}\,{\cal M}_{IJ}\,F_{\nu_1\dots\nu_{p+1}}^I\,F^{\nu_1\dots\nu_{p+1}\,J}
\;,
\qquad
\mbox{etc.}\;,
\label{LkinB}
\eea
with the abelian field strength\footnote{Throughout,
when antisymmetrizing indices $[\mu_1\dots\mu_p]$, 
we use the normalization with total weight one,
i.e.\ 
$X_{[\mu\nu]}=\frac12(X_{\mu\nu}-X_{\nu\mu})$, etc.}
 $F^I_{\nu_1 \dots \nu_{p+1} }=
(p\!+\!1)\,\partial^{\vphantom{I}}_{[\nu_1} B^I_{\nu_2 \dots \nu_p\nu_{p+1}] }$\,.
\smallskip

An important ingredient in the construction of supergravity theories
by dimensional reduction which will also be of relevance in the construction
of the gaugings below is the on-shell duality
between massless $p$-forms and $(D-p-2)$-forms in $D$ space-time
dimensions. This simply reflects the fact that these forms carry the 
same representation
under the little group ${\rm SO}(D\!-\!2)$.
Specifically, it follows from~(\ref{LkinB}) that the field equation 
of a $p$-form $B^I$ to lowest order in the fields
take the form
\bea
\partial^{\mu}\,({\cal M}_{IJ}\,F^J_{\mu\nu_1 \dots \nu_p}) &=& 0
\;.
\label{fieldF}
\eea
The full field equations receive higher-order terms in the fermions as
well as contributions from possible topological terms.
At the same time, the abelian field strength $F^I$ is subject to the Bianchi
identity
\bea
\partial^{\vphantom{J}}_{[\nu_1}\,F^I_{\nu_2 \dots \nu_{p+2}] } &=& 0
\;.
\label{BianchiF}
\eea
In terms of the dual field strength 
\bea
G_{\mu_1 \dots \mu_{D-p-1}\,I} &\equiv& 
\frac{e}{(p\!+\!1)!}\,\varepsilon_{\mu_1 \dots \mu_{D-p-1} \nu_1 \dots \nu_{p+1}}\,
{\cal M}_{IJ}\,F^{\nu_1 \dots \nu_{p+1}\,J }
\;,
\label{duality}
\eea
the equations (\ref{fieldF}) and (\ref{BianchiF}) take the form
\bea
\partial_{[\mu_1}\,G_{\mu_2 \dots \mu_{D-p}]\,I } &=& 0
\;,\qquad
\mbox{and}\quad
\partial^{\mu}\,({\cal M}^{IJ}\,G_{\mu\nu_1 \dots \nu_{D-p-2} \,J}) ~=~ 0
\;,
\eea
respectively, i.e.\ equations of motion and Bianchi identities exchange their roles
and locally we can define the dual $(D-p-2)$-forms $C_I$ by
\bea
G_{\mu_1 \dots\mu_{D-p-1}\,I }&\equiv&
(D\!-\!p\!-\!1)\,\partial_{[\mu_1} C_{\mu_2 \dots\mu_{D-p-1}] \,I}
\;.
\label{dualform}
\eea
Dynamics of the massless $p$-forms $B^I$ can thus equivalently 
be described in terms of their dual $(D\!-\!p\!-\!2)$-forms $C_I$,
transforming in the dual representation under the global symmetry ${\rm G}$. 
This equivalence extends to the full non-linear theory,
i.e.\ in presence of Chern-Simons terms and couplings to
the fermion fields.

As a result, there are in general several different off-shell
formulations of a given ungauged supergravity which are 
on-shell equivalent only after dualizing part of their field content 
according to~(\ref{duality}).
It may not always be possible to get rid of all $p$-forms 
by this dualization,
as the presence of topological terms with explicit appearance of the 
gauge fields can prevent the elimination of these 
fields by virtue of~(\ref{duality}). 
There is however always a version of the theory 
in which all forms are 
dualized to lowest possible degree.
This is the version in which the largest global
symmetry group ${\rm G}$ is manifest,
and the $p$-forms couple with kinetic terms 
(\ref{Lkin}) and (\ref{LkinB}), respectively.
We will see in section~\ref{sec:gauging} that all gaugings of supergravity
can be obtained as deformations of this 
particular version of the ungauged theory.

Let us finally mention that the duality~(\ref{duality}) naturally extends to 
the scalar fields ($p=0$), which are hence on-shell dual to 
$(D-2)$-forms. Due to the non-linear coupling of scalar fields 
discussed in section~\ref{subsec:scalar}, the representation assignment
is slightly different: $(D-2)$-forms generically transform in the full
adjoint representation of the global symmetry group ${\rm G}$
with (\ref{duality}) replaced by
\bea
G_{\mu_1 \dots \mu_{D-1}\,\alpha} &\equiv& 
e\,\varepsilon_{\mu_1 \dots \mu_{D-1} \nu}\,
\,j^{\nu}_\alpha
\;,
\label{duality-scalars}
\eea
where $j^{\nu}_\alpha$ is the conserved Noether-current
associated with the symmetry generated by $t_\alpha$.
Again, this duality cannot be used to eliminate all scalar 
fields (as e.g.\ the scalar dependence of (\ref{LPP}) cannot be expressed
exclusively in terms of the $j^{\nu}_\alpha$),
but only those fields on which the action of ${\rm G}$
is realized as a shift isometry $\phi^i \rightarrow \phi^i+c^i$\,.
The apparent mismatch between the 
$({\rm dim}\,{\rm G}-{\rm dim}\,{\rm K})$ physical scalar fields and 
the number of $(D\!-\!2)$-forms defined in (\ref{duality-scalars})
is explained by the fact that not all the Noether currents $j_\alpha$
are independent: it follows from the structure of the coset space sigma-model 
that ${\cal V}^{-1} (j_\alpha t^\alpha) {\cal V} \in \mathfrak{p}$
for the Noether current associated with~(\ref{LPP}). 
This implies $({\rm dim}\,{\rm K})$ linear constraints on the fields strengths $G_\alpha$.

%%%%%%%%%%%%%%%%%%%%%%%%%%%%%%%%%%%%%%%
\subsection{Self-duality in even dimensions}
\label{subsec:even}
%%%%%%%%%%%%%%%%%%%%%%%%%%%%%%%%%%%%%%%

Employing the on-shell duality~(\ref{duality}) one can always achieve a
formulation of the theory in which all forms are dualized to a degree $p\le [(D\!-\!1)/2]$
and appear with the kinetic terms (\ref{Lkin}), (\ref{LkinB}).
A subtlety arises in even dimensions $D=2K$ 
for the coupling of the $(K\!-\!1)$-forms.
Due to the duality~(\ref{duality}) 
between $(K\!-\!1)$-forms and $(K\!-\!1)$-forms,
these forms 
appear in pairs $(B^\Lambda, B_\Lambda)$
of which only the first half enters the Lagrangian
and carries propagating degrees of freedom while the
other half is defined as their on-shell duals.
For the maximal theories, one observes that only together 
the forms $B^\Lambda$ and their on-shell duals $B_\Lambda$ 
transform in a $2m$-dimensional irreducible linear representation 
$B^P=(B^\Lambda, B_\Lambda)$
of the symmetry group~${\rm G}$, shown in table~\ref{table:pforms}.
As a consequence, in even dimensions ${\rm G}$ is 
only realized as an on-shell symmetry.
E.g.\ the ${\cal N}=8$ supergravity multiplet in $D=4$ dimensions carries 28
vector fields which show up in the Lagrangian, but it is only together
with their 28 magnetic duals that they form the fundamental ${\bf 56}$
representation of ${\rm G}={\rm E}_{7(7)}$~\cite{Cremmer:1979up}.

The analogue of the duality equation~(\ref{duality}) in this case is the
on-shell ${\rm G}$-covariant twisted self-duality equation~\cite{Cremmer:1997ct}
\bea
F^P_{\nu_1\dots\nu_K} &=&
-\frac{e}{K!}\,\varepsilon_{\nu_1\dots\nu_K\mu_1\dots\mu_K}\,
\Omega^{PQ}{\cal M}_{QR}\,
F^{\mu_1\dots\mu_K\,R}
\;,
\label{dualityeven}
\eea
for the ${\rm G}$-covariant abelian field strength
$F^P$, the symmetric matrix ${\cal M}_{PQ}$ from
(\ref{defM}) evaluated in the corresponding 
$2m$-dimensional representation of~${\rm G}$,
and the matrix $\Omega^{PQ}$ given by
\bea
\Omega^{PQ}\equiv
\left(
\begin{array}{cc}
0&{\mathbb{I}}_m\\
\epsilon{\mathbb{I}}_m & 0
\end{array}
\right)
\;,\qquad
\mbox{with}\;\;
\epsilon=(-1)^{K+1}
\;.
\label{omega}
\eea
Consistency of (\ref{dualityeven}) requires that
the matrix $\Omega^{PQ}{\cal M}_{QR}$ squares to $\epsilon$,
such that the total operator acting on $F^P$ on the r.h.s.\ of 
this equation squares to the identity.
This translates into the condition
 \bea
{\cal M}_{IK}\,\Omega^{KL}{\cal M}_{LJ} 
&=&  \Omega_{IJ}
\;,
\label{condM}
\eea
i.e.\ requires the matrix ${\cal M}_{PQ}$
to be symplectic/orthogonal for $K$ even/odd, respectively.
Indeed, table~\ref{table:groups} shows that
the global symmetry groups ${\rm G}$ in even dimensions
can be embedded into ${\rm Sp}(m,m)$ and ${\rm SO}(m,m)$, 
for $K$ even/odd, respectively. For the
non-simple groups appearing in the list
it is sufficient that the factor under which the 
($K-1$)-forms transform non-trivially can be embedded into
${\rm Sp}(m,m)$ or ${\rm SO}(m,m)$, respectively.
E.g.\ for the maximal theory in $D=8$, it is the ${\rm SL}(2)\sim {\rm Sp}(1,1)$ 
factor which mixes three-forms
with their on-shell duals~\cite{Salam:1984ft}.
In $D=6$ dimensions we identify 
${\rm SO}(5,5)$, and
${\rm GL}(1) \sim {\rm SO}(1,1)$, respectively,
while in $D=4$ dimensions, (\ref{condM}) is
ensured by the symplectic embeddings
${\rm E}_{7(7)}\subset {\rm Sp}(28,28)$, and
${\rm SL}(2)\times {\rm SO}(6,6\!+\!n) \subset {\rm Sp}(12\!+\!n,12\!+\!n)$,
respectively.
As $\Omega$ is a group-invariant tensor, 
equation~(\ref{dualityeven})
is manifestly ${\rm G}$-covariant.
Let us mention that for odd $K$ there are theories in which
the eigenvalues of $\Omega^{PQ}{\cal M}_{QR}$ do not come in real
pairs $\pm1$, such that $\Omega$ does not take the form (\ref{omega}).
These theories 
(which include the ten-dimensional IIB 
theory and the six-dimensional
chiral theories with tensor multiplets)
do not admit an action and we shall not
consider them in the rest of this section.

In order to lift equation~(\ref{dualityeven}) to an action, one 
employs the split $B^P=(B^\Lambda, B_\Lambda)$ 
and constructs the action 
in terms of half of the fields $B^\Lambda$ considered as independent propagating 
(electric) fields, 
while the $B_\Lambda$ are defined via (\ref{dualityeven}) as their on-shell 
(magnetic) duals~\cite{Gaillard:1981rj}.
To lowest order in the fields, the proper Lagrangian is given by 
\bea
{\cal L}_{\rm kin} &=&
\ft1{2K!}e\,
{\cal I}_{\Lambda\Sigma}(\phi)\,
F_{\nu_1\dots\nu_{K}}^{\Lambda}
F^{\nu_1\dots\nu_{K}\,\Sigma}
+
\ft1{2(K!)^2}\,
\varepsilon^{\mu_1\dots\mu_K\nu_1\dots\nu_K}\,
{\cal R}_{\Lambda\Sigma}(\phi)\,
F_{\mu_1\dots\mu_{K}}^{\Lambda}
F_{\nu_1\dots\nu_{K}}^{\Sigma}
\;,\nonumber\\
\label{Leven}
\eea
in terms of the $m$ abelian field strengths $F^\Lambda$,
with the kinetic matrices ${\cal I}_{\Lambda\Sigma}(\phi)$ and
${\cal R}_{\Lambda\Sigma}(\phi)$ related to 
the matrix ${\cal M}_{PQ}$ as
\bea
{\cal M}_{PQ} &\equiv&
-\left(
\begin{array}{cc}
{\cal I}-\epsilon{\cal R}{\cal I}^{-1}{\cal R}&\epsilon {\cal R}{\cal I}^{-1}\\
- {\cal I}^{-1}{\cal R} & {\cal I}^{-1}
\end{array}
\right)
\;.
\label{defMeven}
\eea
Indeed, it is easy to verify that an arbitrary symmetric matrix ${\cal M}$ satisfying
(\ref{condM}) can be parametrized as (\ref{defMeven}) in terms of two
matrices ${\cal I}_{\Lambda\Sigma}={\cal I}_{\Sigma\Lambda}$ and
${\cal R}_{\Lambda\Sigma}=-\epsilon{\cal R}_{\Sigma\Lambda}$,
obeying the correct symmetry properties according to their 
appearance in (\ref{Leven}). Moreover, ${\cal I}_{\Lambda\Sigma}$
is negative definite, such that the kinetic term in (\ref{Leven})
comes with the correct sign.

The field equations implied by the Lagrangian~(\ref{Leven})
are conveniently expressed in terms of the
dual field strength defined as
\bea
G_{\mu_1\dots\mu_{K}\,\Lambda} &\equiv&
(-1)^{K+1}\,
\varepsilon_{\mu_1\dots\mu_K\nu_1\dots\nu_K}\,
\frac{\delta{\cal L}}{\delta F^\Lambda_{\nu_1\dots\nu_K}}
\;.
\label{defGeven}
\eea
as
\bea
\partial_{[\mu_1}G_{\mu_2\dots\mu_{K+1}]\,\Lambda} ~=~ 0\;,
\eea
allowing for the introduction of the $m$ dual
$(K\!-\!1)$-forms $B_\Lambda$ according to (\ref{dualform}), 
in terms of which Bianchi identities and field equations exchange 
their roles as in the previous section.
Upon manipulation of~(\ref{defGeven}) 
one recovers the manifestly ${\rm G}$-covariant form of
the field equations~(\ref{dualityeven}) with 
$F^P=(F^\Lambda, G_\Lambda)$.
The action~(\ref{Leven}) can be extended to the full non-linear 
theory, including higher order topological terms and fermionic fields.

Equation~(\ref{defMeven}) shows that the linear action of ${\rm G}$ 
on ${\cal M}$~(\ref{transM}) generically translates into a non-linear action
on the kinetic matrices
${\cal I}$, ${\cal R}$.
Moreover, the action of ${\rm G}$ mixes the components of the vector
$(B^\Lambda, B_\Lambda)$ and thus the forms 
appearing in~(\ref{Leven}) 
with their on-shell duals.
As a consequence, ${\rm G}$ is not a symmetry
of the Lagrangian and only realized on-shell, as is manifest in~(\ref{dualityeven}). 
Only its subgroup
corresponding to triangular generators
\bea
(t_{\alpha})_P{}^Q &=& \left(
\begin{array}{cc}
*&*\\
0&*
\end{array}
\right)
\;,
\eea
is realized as an off-shell symmetry of the Lagrangian.

Different electric/magnetic splits $B^K\rightarrow (B^\Lambda, B_\Lambda)$
correspond to different electric frames and are related by 
symplectic/orthogonal rotation. These give rise to 
different off-shell formulations which are on-shell equivalent.
In particular, the off-shell symmetry group depends on the 
particular choice of the electric frame.

An example of these structures is the half-maximal theory in $D=4$
dimensions with coset space (\ref{half-maximal}) 
whose vector fields $A_\mu^{m\alpha}$ 
transform in the bifundamental representation of ${\rm SO}(6,6+n)\times{\rm SL}(2)$. 
A convenient electric/magnetic split is
$A^{m\alpha}\rightarrow (A^{m+},A^{m-})$ breaking up the 
${\rm SL}(2)$ doublet index $\alpha$.
With the matrix ${\cal M}_{m\alpha,n\beta}$ of (\ref{defM})
factorizing according to
\bea
{\cal M}_{m\alpha,n\beta} &=& 
{\cal M}_{mn}\,{\cal M}_{\alpha\beta} \;,
\eea
into an ${\rm SO}(6,6+n)$ matrix ${\cal M}_{mn}$ and the ${\rm SL}(2)$
matrix ${\cal M}_{\alpha\beta}$ of (\ref{Mex}),
the Lagrangian (\ref{Leven}) is found via (\ref{defMeven}) to be
\bea
{\cal L}_{\rm kin} &=&
-\ft1{4}\,
\Big(
e\,
\Im \tau \,{\cal M}_{mn}\,
F_{\mu\nu}^{m+}
F^{\mu\nu\,n+}
+
\ft1{2}
\varepsilon^{\mu\nu\sigma\tau}\,
\Re\tau\,{\eta}_{mn}\,
F_{\mu\nu}^{m+}
F_{\sigma\tau}^{n+}
\Big)
\;,
\eea
with the ${\rm SO}(6,6+n)$ invariant metric $\eta_{mn}$.
This is an ${\rm SO}(6,6+n)$ covariant electric frame 
in which the ${\rm SL}(2)$ global symmetry is realized only on-shell.
In other frames, the full ${\rm SL}(2)$ may be elevated to an off-shell
symmetry, but only a ${\rm GL}(6)$ subgroup of ${\rm SO}(6,6+n)$
remains realized off-shell.

Let us finally mention that the case of $D=2$ supergravity is particularly subtle,
as the self-duality of forms discussed in this section applies to the scalar
fields of the theory.
The formalism thus needs to be merged with the non-linear 
realization of the scalar isometries discussed in section~\ref{subsec:scalar}.
As a result, the duality between scalar fields is not of the simple  
type as for the $p$-forms, but rather leads to an infinite chain of mutually
dual scalar fields, on which the infinite-dimensional global symmetry group~${\rm G}$
can be linearly realized. See~\cite{Julia:1981wc,Breitenlohner:1986um,Nicolai:1987kz,Nicolai:1998gi,Bernard:1997et} for details.

%%%%%%%%%%%%%%%%%%%%%%%%%%%%%%%%%%%%%%%
%%%%%%%%%%%%%%%%%%%%%%%%%%%%%%%%%%%%%%%

\section{Gauging Supergravity --- Covariant Formulation}
\label{sec:gauging}

%%%%%%%%%%%%%%%%%%%%%%%%%%%%%%%%%%%%%%%
%%%%%%%%%%%%%%%%%%%%%%%%%%%%%%%%%%%%%%%

In the last section, we have reviewed how the field content and the action
of ungauged supergravity are organized by the global symmetry group ${\rm G}$.
Scalar fields and $p$-form fields transform in a non-linear 
and in linear representations of ${\rm G}$, respectively.
We will now discuss the gaugings of the theory. I.e.\
according to the general discussion of section~\ref{sec:intro}
we will select a subgroup 
${\rm G}_0\subset{\rm G}$ and promote it to a local symmetry. 
This can be considered as a deformation of the ungauged theory 
and we shall discuss which additional couplings have to be
imposed along the way.
We will employ the covariant formalism 
of~\cite{Nicolai:2000sc,Nicolai:2001sv,deWit:2002vt,deWit:2004nw,deWit:2005hv} in which the gaugings are encoded in the embedding tensor
which may be characterized group-theoretically.

%%%%%%%%%%%%%%%%%%%%%%%%%%%%%%%%%%%%%%%
\subsection{The embedding tensor}
\label{subsec:embedding}
%%%%%%%%%%%%%%%%%%%%%%%%%%%%%%%%%%%%%%%

As we have reviewed in the last section, under the 
non-abelian global symmetry group~${\rm G}$,
the bosonic fields of ungauged supergravity transform as
\bea
\delta\,{\cal V}&=& \Lambda^{\alpha}\,t_{\alpha}\,{\cal V}\;,
\non
\delta A_{\mu}^{M} &=& 
-\Lambda^{\alpha}\,(t_{\alpha})_{N}{}^{M}\,A_{\mu}^{N}
\;,
\non[.3ex]
\mbox{etc.}
\;,
\label{actionungauged}
\eea
with constant parameters $\Lambda^\alpha$, $\alpha=1, \dots, {\rm dim}\,{\rm G}$. 
In addition, the $n_{\rm v}$
vector fields in the theory possess the standard abelian gauge symmetry
${\rm U}(1)^{n_{\rm v}}$\,:
\bea
\delta\,A_{\mu}^{M} &=& \partial_{\mu}\,\Lambda^{M}\;,
\eea
with coordinate-dependent parameters $\Lambda^M=\Lambda^M(x)$.
Similarly, higher-rank $p$-forms appear with the corresponding
abelian tensor gauge symmetry.

Gauging corresponds to promoting a subgroup ${\rm G}_0\subset{\rm G}$
to a local symmetry. This subgroup can be defined by 
selecting a subset of generators within 
the global symmetry algebra $\mathfrak{g}={\rm Lie}\,{\rm G}$.
Denoting these generators by $X_M$, the associated symmetries
can be made local by introducing standard covariant derivatives according to
\bea
\partial_\mu &\longrightarrow& D_\mu ~\equiv~
\partial_\mu - gA_\mu^M\,X_M
\;,
\label{covariant}
\eea
where we also introduce the gauge coupling constant $g$. 
A general set of $n_{\rm v}$ generators in 
$\mathfrak{g}$ can be described as
\bea
X_{M} &\equiv& \Theta_{M}{}^{\alpha}\,t_{\alpha}~\in~\mathfrak{g}
\;,
\label{generators}
\eea
by means of a constant tensor $\Theta_{M}{}^{\alpha}$,
the {\em embedding tensor}, which describes the explicit embedding of the
gauge group ${\rm G}_0$ into the global symmetry group ${\rm G}$.
For the moment, we can simply consider this object as a constant 
$(n_{\rm v} \times {\rm dim}\,{\rm G})$ matrix with its two indices $M$ and $\alpha$
in a fundamental and the adjoint representation of ${\rm G}$,
respectively.
It combines the full set of deformation parameters.
The dimension of the gauge group is given by
the rank of the matrix $\Theta_{M}{}^{\alpha}$.

The advantage of explicitly parametrizing the gauge group generators 
as in~(\ref{generators}) is, that this allows to keep the entire construction 
formally ${\rm G}$-covariant.
As it will turn out, the gauging can be entirely parametrized 
in terms of the embedding tensor~$\Theta_{M}{}^{\alpha}$.
The deformed equations of motion remain manifestly ${\rm G}$-covariant
if the embedding tensor is treated as a
spurionic object that simultaneously transforms under ${\rm G}$ 
according to the structure of its indices.
It is only upon specifying a particular choice for $\Theta_{M}{}^{\alpha}$ that 
we select a particular gauge group~${\rm G}_0$, and the 
global symmetry~${\rm G}$ is broken.
The embedding tensor will always appear together with the 
coupling constant $g$ we have introduced in (\ref{covariant}).
The latter could thus be absorbed by rescaling $\Theta_{M}{}^{\alpha}$,
but we will keep it in the following for book-keeping purpose.

Having introduced the covariant derivatives, the theory should be
invariant under the standard combined transformations
\bea
\delta\,{\cal V}&=& g\Lambda^{M}X_{M}\,{\cal V}\;,
\non
\delta A_{\mu}^{M} &=&  
  \partial_\mu \Lambda^{M} + g A_\mu^{N}
  X_{NP}{}^{M} \Lambda^{P} ~=~ 
  D_\mu\Lambda^{M} 
\;,
\label{gauge}
\eea
with local parameter $\Lambda^{M}=\Lambda^{M}(x)$ and 
$X_{NK}{}^{M}\equiv\Theta_{N}{}^{\alpha}\,(t_{\alpha})_{K}{}^{M}$.
This is of course not true for an arbitrary choice 
of $\Theta$. 
In particular, consistency requires
that the generators~(\ref{generators})
close into a subalgebra of $\mathfrak{g}$.
This in turn translates into a set of non-trivial constraints
on the embedding tensor.
In the following, we shall work out the complete set
of constraints which $\Theta$ must satisfy
in order to achieve a theory with local gauge invariance~(\ref{gauge}).
As it turns out, there are in general two sets of constraints,
a quadratic and a linear one.
They can be formulated as ${\rm G}$-covariant homogeneous
equations in $\Theta$ which allows to
construct solutions by purely group-theoretical methods.
Eventually, every solution to this set of constraints
will give rise to a consistent Lagrangian with local gauge symmetry~(\ref{gauge}).

The first set of constraints is bilinear in $\Theta$ and
very generic. It states that the tensor~$\Theta$ is invariant under
the action of the generators~(\ref{generators}) of the local
gauge symmetry. 
Note that $\Theta$ is almost never a ${\rm G}$-invariant tensor,
as follows already from the different nature of its two indices
(except in $D=3$ dimensions, where vector fields transform in the
adjoint representation of ${\rm G}$).
Consistency of the gauged theory however requires that $\Theta$
must be invariant under the action of the subgroup ${\rm G}_0$. 
As this subgroup is precisely defined by projection with $\Theta$, 
together this leads to a {\em quadratic} constraint in~$\Theta$:
\bea
 0 ~\, \stackrel{!}= \,~
 {\cal Q}_{PM}{}^\alpha ~\equiv~ \delta_{ P} \,\Theta_{ M}{}^\alpha 
 &\equiv&
  \Theta_{ P}{}^\beta\,  \delta_{\beta} \,\Theta_{ M}{}^\alpha \non
  &=& \Theta_{ P}{}^\beta (t_\beta)_{M}{}^{ N}
    \Theta_{ N}{}^\alpha +
    \Theta_{ P}{}^\beta f_{\beta\gamma}{}^\alpha 
    \Theta_{ M}{}^\gamma  \;,
\label{quadraticconstraint}
\eea
where we have used the fact that
the generators in the adjoint representation are given 
in terms of the structure constants as
$(t_\alpha)_\beta{}^\gamma=-f_{\alpha\beta}{}^\gamma$\,.
Contracting this result with a generator $t_\alpha$, we obtain
the equivalent form
\bea
{[X_{M},X_{N}]} &=&
-X_{MN}{}^{P}\,X_{P}    
\;,\qquad
{\rm with}
\quad
X_{MN}{}^{P}=\Theta_M{}^\alpha (t_\alpha)_N{}^P
\;. 
\label{closure}
\eea
Hence, the gauge invariance of the embedding tensor in particular implies
the closure of the generators~(\ref{generators})
into an algebra. Let us stress however, that the constraint~(\ref{quadraticconstraint})
 is in general stronger than simple closure:
equation (\ref{closure}) in particular implies a non-trivial 
relation upon symmetrization in $(MN)$ 
(upon which the l.h.s.\ trivially vanishes, but the r.h.s.\ does not)
which clearly goes beyond closure.
Nevertheless this condition turns out to be indispensable
and we will come back to it in the next subsection.

Apart from the quadratic constraint~(\ref{quadraticconstraint}),
$\Theta$ must in general satisfy another {\em linear} constraint
which is implied by supersymmetry.
Recall that eventually we wish to construct a theory that does not
only possess the local gauge invariance~(\ref{gauge}) but
also should still be invariant under (a possible deformation of)
supersymmetry. This puts further constraints on $\Theta$,
whose specific form however will in particular depend
on the number of space-time dimensions
and supercharges considered.
Interestingly enough, in many cases 
the linear constraint can already be deduced at a
much earlier stage by purely bosonic considerations
related to consistency of the deformed tensor gauge algebra
that we discuss in the next section.

Let us consider as an example the maximal ${\cal N}=8$, $D=4$ theory.
Its global symmetry group ${\rm G}={\rm E}_{7(7)}$ has 133 generators
while the vector fields transform in the fundamental ${\bf 56}$
representation.
According to its index structure, the embedding tensor $\Theta_M{}^\alpha$ 
thus a priori lives in the tensor product of the fundamental and the
adjoint representation, which decomposes according 
to\footnote{All the tensor products and branchings of representations used
in these lectures can
be found in the appendix of~\cite{Slansky:1981yr} or calculated
with the help of the computer algebra package LiE~\cite{LiE}.}
\bea
\Theta_M{}^\alpha\,:\qquad
{\bf 56} \otimes {\bf 133} &=& 
{\bf 56} \oplus {\bf {912}} \oplus {\bf 6480}
\;.
\label{912}
\eea
Compatibility of the deformation with supersymmetry
can be expressed in an ${\rm E}_{7(7)}$-covariant way and 
restricts the embedding tensor to the ${\bf {912}}$ 
representation in this decomposition~\cite{deWit:2002vt,deWit:2007mt}.
I.e.\ as a matrix $\Theta_M{}^\alpha$ has only 912 linearly
independent entries.
We will sketch the argument in 
section~\ref{subsec:Lagrangian} below.

It is interesting to note that in fact also the 
${\bf 56}$ part of the embedding tensor in the decomposition~(\ref{912})
defines a consistent deformation which however 
requires simultaneous gauging
of the $\mathbb{R}^+$ on-shell conformal rescaling symmetry of
$D=4$ supergravity, see~\cite{trombone} for details. 
As a result, the corresponding
theory no longer admits an action but can be constructed as
a supersymmetric deformation of the equations of motion.
Deformations of this type have first been 
constructed in ten space-time 
dimensions in~\cite{Howe:1997qt,Lavrinenko:1997qa}.

As another example, we may consider the half-maximal theory
in $D=4$ dimensions, whose scalar fields parametrize the 
coset space~(\ref{half-maximal}) with the vector fields
transforming in the bifundamental $({\bf 2},\!\tinyyoung{\cr}\, )$ 
representation\footnote{The box `$\tinyyoung{\cr}$\,'
here refers to the vector representation of ${\rm SO}(6,6\!+\!n)$
and we use the standard Young tableaux notation for
the higher irreducible representations.
} of 
${\rm SL}(2)\otimes{\rm SO}(6,6\!+\!n)$.
In analogy to (\ref{912}), the embedding tensor a priori
transforms in the tensor product of fundamental
and adjoint representation, which decomposes according to
\bea
({\bf 2},\tinyyoung{\cr}\,) \otimes
\Big(({\bf 3}, 1) + ({\bf 1}, \tinyyoung{\cr \cr} \,\,)\Big)
&=&
2\cdot ({\bf 2},\tinyyoung{\cr}\, ) 
 \oplus({\bf 2},\tinyyoung{\cr \cr \cr}\,\, )
 \oplus({\bf 2}, \tinyyoung{& \cr \cr}\,\, )
 \oplus ({\bf 4},\tinyyoung{\cr}\,)
\;.
\label{13}
\eea
Supersymmetry restricts the embedding tensor to
$ ({\bf 2},\!\tinyyoung{\cr}\, )+({\bf 2},\!\tinyyoung{\cr \cr \cr}\,\, )$,
i.e.\ forbids the last two contributions in (\ref{13}) and poses
a linear constraint among the two terms in the
$({\bf 2},\tinyyoung{\cr}\, )$ representation~\cite{Schon:2006kz}.

In general, the embedding tensor lives within the tensor 
product
\bea
\Theta_M{}^\alpha\,:\qquad
{\cal R}_{{\rm v}^*} \otimes {\cal R}_{\rm adj} &=&
{\cal R}_{{\rm v}^*} \oplus~ \dots
\;,
\label{RR}
\eea
where by ${\cal R}_{{\rm v}^*}$ we denote the representation
dual to the representation ${\cal R}_{{\rm v}}$ in which the vector fields transform,
and the precise form of the r.h.s.\ depends on the particular
group and representations considered.
The linear representation constraint then schematically takes the 
form
\bea
\mathbb{P}\,\Theta &=& 0\;,
\label{linearconstraint}
\eea
and restricts $\Theta$ to some of the representations
appearing on the r.h.s.\ of (\ref{RR}).
The resulting representations for the embedding tensor 
in the maximal supergravities
are collected in table~\ref{table:embedding}.
For the half-maximal theories, the structure is very similar,
however the embedding tensor generically contains
several different irreducible parts, see 
e.g.~\cite{Nicolai:2001ac,Schon:2006kz,Weidner:2006rp,Bergshoeff:2007vb}.

%%%%%%%%%%%%%%%%%%%%%%%%%%%%
\begin{table}[tb]
\begin{center}
\begin{tabular}{r||c||c|c||c|}
$D$ & ${\rm G}$ & ${\cal R}_{\rm adj}$ & ${\cal R}_{\rm v}$ & $\Theta_M{}^\alpha$      \\
\hline
9   & ${\rm GL}(2)$  & 
${\bf 1}^0 + {\bf 3}^0$ & ${\bf 1}^{-4} + {\bf 2}^{+3}$ & ${\bf 2}^{-3}+{\bf 3}^{+4}$ 
\\
8   & ${\rm SL}(2)\! \times \!{\rm SL}(3)$  & $({\bf 3},{\bf 1})\!+\!({\bf 1},{\bf 8})$
& $({\bf 2},{\bf 3}')$ &$({\bf 2},\!{\bf 3})\!+\!({\bf 2},\!{\bf 6}')$\\
7   & ${\rm SL}(5)$   &  ${\bf 24}$ & ${\bf 10}'$ & ${\bf 15}+{\bf 40}'$ \\
6   &  ${\rm SO}(5,5)$  &  ${\bf 45}$ & ${\bf 16}_c$ & ${\bf 144}_c$ \\
5   & E$_{6(+6)}$   &  ${\bf 78}$ & ${\bf 27}'$ & ${\bf 351}'$ \\
4   & E$_{7(+7)}$  &  ${\bf 133}$ & ${\bf 56}$ & ${\bf 912}$  \\
3   & E$_{8(+8)}$  & ${\bf 248}$ & ${\bf 248}$ & ${\bf 1}\!+\!{\bf 3875}$ \\
2   & E$_{9(+9)}$  & $\;\;{\bf\cal R}_{\rm adj}$ & ${\bf \Lambda}_1$ & 
${\bf \Lambda}_{1*}$ 
\end{tabular}
\end{center}
\caption{\small
Embedding tensor $\Theta_M{}^\alpha$ in maximal supergravity.
In $D=2$ dimensions, ${\bf\cal R}_{\rm adj}$ and ${\bf \Lambda}_1$
refer to the infinite-dimensional adjoint and basic representation
of the affine algebra ${\rm E}_{9(9)}=\widehat{{\rm E}}_{8(8)}$, respectively.
}\label{table:embedding}
\end{table}

As we have already mentioned above,
the linear representation constraint
follows from consistency of the deformation
with supersymmetry in first order of $\Theta$,
see section~\ref{subsec:Lagrangian} below.
In many cases however, this constraint can already 
be deduced from purely bosonic considerations 
and we will see examples of
this in the following.
E.g.\ in $D=4$ dimensions, it can be shown~\cite{deWit:2005ub} 
that the embedding
tensor must in general satisfy the linear constraint
\bea
X_{(MN}{}^P\,\Omega_{K)P}&=&0 
\;,
\label{linear4D}
\eea
with $X_{MN}{}^P$ from (\ref{closure}) and the symplectic matrix 
$\Omega_{KP}$ from (\ref{omega}),
in order to achieve a bosonic Lagrangian with the local
symmetry~(\ref{gauge}).
This condition does not make any reference to supersymmetry
but in particular reproduces the constraints 
given above for the ${\cal N}=8$ and the ${\cal N}=4$
theory, respectively.\footnote{It has recently been shown that 
in ${\cal N}=1$ theories this condition may be replaced by an 
inhomogeneous equation in $\Theta$ in order to cancel
the quantum anomaly cubic in $\Theta$~\cite{DeRydt:2008hw}. Here,
we restrict the discussion to the classical theories.
}
\smallskip

To summarize the discussion of this section, any gauging of the theory is entirely
encoded in the choice of the embedding tensor $\Theta_M{}^\alpha$
which according to (\ref{generators}) defines the embedding of
the gauge group ${\rm G}_0$ into the global symmetry group ${\rm G}$
and via (\ref{covariant}) defines the new minimal couplings of vector
fields to the remaining matter fields.
Consistency of the deformation is expressed by a set of algebraic constraints
quadratic~(\ref{quadraticconstraint}) and linear~(\ref{linearconstraint})
in the embedding tensor, respectively.
While it is rather straightforward to verify
that the constraints are necessary, it has to be checked case by case 
(i.e.\ for the various dimensions $D$ and number of supercharges $N$) that
indeed they are sufficient to define a consistent gauging.
In particular, the action of the gauged theory
${\cal L}_{\rm gauged}$ must be constructed separately in the
various space-time dimensions. However once this action has been worked
out for generic $\Theta$, all particular gaugings are straightforwardly obtained
as specific choices of~$\Theta$.\footnote{We should mention that
for theories with low number of supersymmetries additional quadratic
constraints beyond~(\ref{quadraticconstraint}) may arise,
such as the locality of electric and magnetic charges
in four dimensions~\cite{deWit:2005ub}. 
For the maximal and half-maximal theories
discussed here, the quadratic constraint~(\ref{quadraticconstraint})
is a sufficient condition.}

The classification of the possible gaugings in a given space-time 
dimension thus reduces to the analysis of simultaneous solutions of 
the constraints (\ref{quadraticconstraint}) and~(\ref{linearconstraint}).
While the latter can be directly solved by working out the explicit projection,
the quadratic constraint is in general difficult to solve
and does not possess a solution in closed form.
The counting of inequivalent gaugings 
in general remains an unsolved problem.
The strategy we will pursue in the following in order to 
identify the gaugings relevant for the description of
particular flux compactifications is the following:
\begin{itemize}
\item
work out in $D$ dimensions the universal gauged Lagrangian,
i.e.\ the deformation for a generic $\Theta$
solving the constraints 
(\ref{quadraticconstraint}) and~(\ref{linearconstraint}).
\item
identify (exploiting the symmetries related to 
the higher-dimensional origin) among the components
of $\Theta$ those corresponding to
particular flux parameters.
\item
evaluate the quadratic constraint and the 
general formulas for this particular choice of $\Theta$.
\item
evaluate the universal formulas for the Lagrangian and in particular the
scalar potential for this particular choice of $\Theta$.
\item
moreover, working out the action of ${\rm G}$ on $\Theta$ 
allows to directly determine
the transformation of the flux parameters under the 
duality group.
\end{itemize}
We shall illustrate this procedure
in section~\ref{sec:flux} with several explicit examples.

%%%%%%%%%%%%%%%%%%%%%%%%%%%%%%%%%%%%%%%
\subsection{Deformed tensor gauge algebra}
%%%%%%%%%%%%%%%%%%%%%%%%%%%%%%%%%%%%%%%

We have in the last section defined a gauging
in terms of the embedding tensor $\Theta_M{}^\alpha$
and started to render the theory invariant under the
local symmetry (\ref{gauge})
by introducing covariant derivatives~(\ref{covariant}).
Apart from the minimal couplings induced by~(\ref{covariant}),
the field strengths of the vector fields need to be modified
in order to capture the non-abelian nature of the new
gauge group. 
This will lead us to a deformation of the higher-rank tensor gauge 
algebra, intertwining $p$-forms and ($p\!+\!1)$-forms.
The natural ansatz for the non-abelian 
field strength of the vector fields is
\bea
  {\cal F}_{\mu\nu}^{M} &=& \partial_\mu A_\nu^{M}
  -\partial_\nu
  A_\mu^{M} + g X_{[{NP}]}{}^{M}
  \,A_\mu^{N} A_\nu^{P} \;, 
\label{defF}
\eea
but we shall see in the following that this is actually not sufficient.

Let us start from the gauge algebra~(\ref{closure})
\bea
{[X_{M},X_{N}]} &=&
-X_{MN}{}^{P}\,X_{P}    
\;. 
\label{closure2}
\eea
with the ``structure constants''
\bea
  X_{ M  N}{}^{ P} &\equiv&
  \Theta_{M}{}^\alpha\,(t_\alpha)_{ N}{}^{ P} 
 ~\equiv~ X_{[ M  N]}{}^{ P} 
  +  Z^{P}{}_{{MN}}\;,
  \label{defZ}
\eea
where we have introduced the notation
$Z^{P}{}_{{MN}} \equiv X_{({MN})}{}^{P}$ for the 
symmetric part in $X_{{MN}}{}^{P}$ which is 
generically non-vanishing.
E.g.\ for the maximal theories with
irreducible embedding tensor $\Theta_{M}{}^\alpha$
(i.e.\ for $D=4, 5, 6$, cf.~table~\ref{table:embedding}),
one may show that there is no gauging for which $Z^{P}{}_{{MN}}$ vanishes.

As the l.h.s.\ of (\ref{closure2}) is manifestly antisymmetric in $[MN]$,
so must be the r.h.s., hence
we can deduce that $Z^{P}{}_{{MN}}$ vanishes upon 
contraction with another generator
\bea
Z^{P}{}_{{MN}}\,X_P &=& 0 \;,
\label{ZX}
\eea
as a direct consequence of the quadratic constraint~(\ref{closure2}).
The presence of a symmetric part in the ``structure constants'' $X_{MN}{}^{ P}$
might seem like a bagatelle and simply motivate the definition of
the explicitly antisymmetrized $X_{[MN]}{}^{ P}$ as the true structure constants.
However, these latter objects fail to satisfy the Jacobi identities:
\bea
 {X_{[MN]}{}^P\, X_{[QP]}{}^R + X_{[QM]}{}^P\, X_{[NP]}{}^R  + X_{[NQ]}{}^P \,
 X_{[MP]}{}^R} = - Z^{R}{}_{P[Q}\, X_{MN]}{}^P \;,  
 \label{nonJac}
\eea
the violation being again proportional to the tensor $Z^{P}{}_{{MN}}$.
Again the standard Jacobi identity is satisfied 
upon further contraction with a generator $X_R$ 
as a consequence of (\ref{ZX}).
This is certainly enough for consistency of the algebra~(\ref{closure2}).
However, it shows up as a problem in the definition of a 
suitable covariant field strength tensor 
which is needed in order to construct a proper covariantization of the
abelian kinetic term~(\ref{Lkin}), a covariant coupling to the
fermionic fields, etc.

Namely, as a consequence of (\ref{nonJac})
the standard non-abelian field strength~(\ref{defF})
turns out to be not fully covariant. Under the
new gauge transformations~(\ref{gauge}) it transforms as
\bea
  \delta {\cal F}_{\mu\nu}^{M}&=& -g\, \Lambda^{P}
  X_{PN}{}^{M}
  \,{\cal F}_{\mu\nu}^{N} + 2 g\, Z^{M}{}_{PQ}
  \Big( \Lambda^{P}{\cal F}_{\mu\nu}^{Q} -\,A_{[\mu}^{P}\,\delta A_{\nu]}^{Q} \Big) \;, 
  \label{noncov}
\eea 
of which only the first term would correspond to a standard homogeneous
covariant transformation. Note however that
all the unwanted terms appear contracted with the tensor $Z^{M}{}_{PQ}$
and thus vanish in absence of $Z^{M}{}_{PQ}$.
In particular, together with (\ref{ZX}) 
this implies that the combination ${\cal F}_{\mu\nu}^{M}\,X_M$,
which e.g.\ shows up in the commutator of covariant derivatives as
\bea
[D_\mu,D_\nu]&=& - g\, {\cal F}_{\mu\nu}^{M}\,X_{M}
\;,
\eea
is a good covariant object. On the other hand, conceivable covariant 
kinetic terms constructed from this object such as
${\rm Tr}\,[{\cal F}_{\mu\nu}^{M}X_{M}\;{\cal F}^{\mu\nu\,N}X_{N}]$
are not smooth deformations of~(\ref{Lkin}).
A priori it thus remains unclear how to properly covariantize
this kinetic term. A similar problem arises in the covariantization
of the kinetic terms for the higher-rank $p$-forms~(\ref{LkinB}). 

The problem we are facing in this construction is 
the price to pay for staying ${\rm G}$-covariant. We have chosen a
somewhat redundant description of the new gauge group in terms
of $n_{\rm v}$ generators~(\ref{generators}), whereas in fact in 
most cases the dimension of the gauge group ${\rm G}_0$ will be strictly smaller
than $n_{\rm v}$. In other words, the matrix $\Theta_M{}^\alpha$ in
general does not have maximal rank, thus not all $X_M$ are linearly 
independent.
Accordingly, the vector fields $A_\mu^M $ split into
\bea
A_\mu^M \;\;
\longrightarrow\;
\left\{
\begin{array}{ll}
A_\mu^m&\!\!\rightarrow\,\,
\mbox{transforming in the adjoint of ${\rm G}_0$}\;,
\\[.8ex]
A_\mu^i &\!\!\rightarrow\,\,
\mbox{remaining vectors, transforming in some rep.\ of ${\rm G}_0$}\;.
\end{array}
\right.
\label{split}
\eea
This clearly poses a problem if the $A_\mu^i$ transform in a non-trivial representation
of ${\rm G}_0$ in which case we cannot write down a consistent gauge theory.
As a consequence, w.r.t.\ this splitting,
 $Z^m{}_{PQ}$ vanishes whereas $Z^i{}_{PQ}\not=0$
creates the problems manifest in equation~(\ref{noncov}).
This is a generic problem in gauged supergravities 
which has been encountered 
ever since such theories have been constructed. Let us briefly discuss how
it has been circumvented in some of the early constructions.
\begin{itemize}

\item
The four-dimensional maximal ${\rm SO}(8)$ gauged theory constructed 
in~\cite{deWit:1982ig} involves only 28 out of the 56 vector fields that
form an irreducible representation of the global symmetry group~${\rm E}_{7(7)}$.
As discussed in section~\ref{subsec:even} above, 
only 28 of these vector fields 
are present in the ungauged action and it is precisely this set which is coupled
to the ${\rm SO}(8)$ generators.
The same situation occurs for the non-compact ${\rm SO}(p,q)$ gaugings
constructed subsequently in~\cite{Hull:1984vg}.
In other words, the vector fields $A_\mu^i$ of (\ref{split}) 
which do not participate
in the gauging all live in the magnetic sector and do not 
appear in the action. It is therefore not a problem that their 
field strengths are not covariant objects.
This distinctive feature of four space-time dimensions has allowed 
the construction of maximal gaugings without having to address 
the problem of non-covariance encountered above.
However, it strikes back in the presence of magnetic charges and tensor 
fields~\cite{DAuria:2004yi}, and we present its covariant solution below.

\item
In five space-time dimensions, the maximal ${\rm SO}(6)$ gauging
constructed in~\cite{Gunaydin:1985cu} involves only 15 out of the 27
vector fields, with the 12 remaining ones 
transforming in a non-trivial representation
of ${\rm SO}(6)$, implying a non-vanishing $Z^i{}_{PQ}$.
The solution found in~\cite{Gunaydin:1985cu} 
corresponds to dualizing
these unwanted vectors into two-forms 
(as a particular application of the dualization
of $p$-forms~(\ref{duality})) which upon gauging turn into 
massive self-dual two-forms, i.e.\ acquire mass terms of the 
type~\cite{Townsend:1983xs}
\bea
\partial^\rho F_{\rho\mu\nu} 
&=& 
e\varepsilon_{\mu\nu\rho\sigma\tau}\,m\,F^{\rho\sigma\tau}
\;,
\label{massive-selfdual}
\eea
with masses $m$ proportional to the inverse coupling constant $g^{-1}$.
As a consequence, these forms continue to carry no more than the three
degrees of freedom of a massless vector field in five dimensions,
consistently keeping the balance of degrees of freedom upon gauging.
The Lagrangian of the gauged theory thus carries only 15
vector fields $A_\mu^m$ with truly covariant field strengths
and 12 massive self-dual two-forms.
The same mechanism has been successfully applied
to other five-dimensional theories, see 
e.g.~\cite{DallAgata:2001vb}.

\item
In seven space-time dimensions the first gaugings of the maximal theory
were constructed in~\cite{Pernici:1984xx} and exhibit a somewhat
particular situation: the entire set of 10 vector fields present in seven 
dimensions is needed to gauge an ${\rm SO}(5)$ group 
(or non-compact versions thereof).
This is related to the fact, that in seven dimensions the embedding tensor
is reducible (cf.~table~\ref{table:embedding}); for gaugings defined by
an embedding tensor in the ${\bf 15}$ of ${\rm SL}(5)$ the tensor 
$Z^{M}{}_{{PQ}}$ vanishes identically.
For these gaugings however a similar problem shows up upon trying to define
a proper covariant field strength for the two-forms. 
In complete analogy to (\ref{noncov}), the naive covariantization of their
abelian field strengths in general does not transform covariantly.
In~\cite{Pernici:1984xx} this problem was circumvented by dualizing
all two-forms into three-forms which upon gauging become self-dual 
massive, analogous to~(\ref{massive-selfdual})
thereby conserving the number of degrees of freedom.
The Lagrangian of the gauged theory of~\cite{Pernici:1984xx}
thus carries only vector fields 
and massive self-dual three-forms.

\end{itemize}

In all these cases it has thus eventually been possible to 
eliminate the extra vector fields $A_\mu^i$ of (\ref{split}) 
from the Lagrangian thereby circumventing the problem of 
the non-covariant field strengths~(\ref{noncov}).
This procedure however requires an explicit breaking of the 
${\rm G}$-covariance: the explicit split (\ref{split}) 
and thus the field content of the gauged theory
depends on the particular gauge group chosen. 
Furthermore, it remains unclear how to proceed in other
space-time dimensions.
E.g.\ a gauging of smaller groups in $D=7$ dimensions cannot
be achieved with this construction. 
Likewise, in even dimensions
there is no analogue of the massive self-duality~(\ref{massive-selfdual})
which was crucial for the correct balance of degrees of freedom.
All this motivates the covariant construction that we shall
present in the following, which appears particularly natural
in the context of flux compactifications.

The covariant ansatz makes use of the fact that the non-covariant 
terms in~(\ref{noncov}) appear projected with the tensor $Z^{M}{}_{PQ}$ 
and defines the full covariant field strengths as~\cite{deWit:2004nw,deWit:2005hv}
\bea
{\cal H}_{\mu\nu}^{{M}} &=& 
{\cal F}_{\mu\nu}^{M}  + g Z^{M}{}_{PQ} 
\,B_{\mu\nu}^{PQ}\;,
\label{covH}
\eea
upon the introduction of two-form tensor fields 
of the type $B_{\mu\nu}^{MN}=B_{[\mu\nu]}^{(MN)}$.
The non-covariant terms in~(\ref{noncov})  can then be absorbed by
postulating the corresponding transformation laws for the two-form fields.
Explicitly, the new field strength ${\cal H}_{\mu\nu}^{{M}}$ 
transforms covariantly under the combined set of gauge transformations
\bea
\delta A_\mu^{M} &=&  D_\mu\Lambda^{M} -
g\,Z^{M}{}_{{PQ}}\,\Xi_\mu^{{PQ}} \;,
\non[.6ex]
\delta B_{\mu\nu}^{MN} &=&
    2\,D_{[\mu}\Xi_{\nu]}^{{MN}} -2
  \Lambda^{{(M}}{\cal H}_{\mu\nu}^{{N)}}
  + 2 \, A_{[\mu}^{{(M}}
  \delta A_{\nu]}^{{N)}}
  \;,
  \label{gaugeAB}
\eea
where $\Xi_\mu^{{MN}}$ labels the tensor gauge
transformations associated with the two-forms.
A non-vanishing tensor $Z^{M}{}_{{PQ}}$ thus induces a
St\"uckelberg-type coupling between vector fields and 
antisymmetric two-forms, as is familiar from massive deformations
of supergravities, e.g.~\cite{Romans:1985tz}.
It is the strength of the covariant formalism to treat all 
possible deformations
(gauged and massive supergravities) on the same footing.

Of course, the two-forms $B_{\mu\nu}^{MN}$ introduced in (\ref{covH})
cannot simply be added to the fields of the theory, as the number
of degrees of freedom is in general carefully balanced by supersymmetry.
Rather, these must be (a subset of) the two-forms that are already present
in the ungauged supergravity.
Their index structure in (\ref{covH}) shows that they 
generically appear in a representation of ${\rm G}$
that is contained in the symmetric tensor product 
$({\cal R}_{\rm v} \otimes {\cal R}_{\rm v})_{\rm sym}$.
Their precise representation can be inferred from inspection of the 
tensor $Z^{M}{}_{PQ}$ under which they appear. 
In turn, this severely constrains the tensor $Z^{M}{}_{PQ}$
which in its indices $(PQ)$ should project only onto those representations
filled by the two-forms in the ungauged theory.
As $Z$ is a function of the embedding tensor~$\Theta$, this eventually leads to a
linear representation constraint of the type~(\ref{linearconstraint}) on $\Theta$.

As an example, let us consider
the case of $D=4$ dimensions where\footnote{
In these formulas we have been raising and lowering 
indices $M, N$ with the symplectic matrix $\Omega^{MN}$
and north-west south-east conventions, i.e. 
$\Theta^{M\alpha}=\Omega^{MN}\Theta_N{}^\alpha$, etc.}
\bea
Z^{K}{}_{MN}&=& X_{(MN)}{}^{K}
\nonumber\\[.5ex]
&=&\ft12 \Theta_{M}{}^{\alpha}\,(t_{\alpha})_{N}{}^{K}
+\ft12 \Theta_{N}{}^{\alpha}\,(t_{\alpha})_{M}{}^{K}
\nonumber\\[.5ex]
&=& -\ft12\Theta^{K}{}^{\alpha}\,(t_{\alpha})_{MN}
+\ft32X_{(MNL)}{}\,\Omega^{KL}
\;,
\eea
where we have made use of the fact that 
following the discussion of section~\ref{subsec:even},
in $D=4$ dimensions the symmetry generators are 
embedded into the symplectic group, i.e.\ 
$(t_{\alpha})_{[MN]}=0$.
Plugging this expression into the covariant field strength~(\ref{covH})
shows that 
if the embedding tensor satisfies the linear constraint $X_{(MNL)}=0$
of~(\ref{linear4D}),
the two-forms in~(\ref{covH}) appear always under the particular projection
\bea
Z^{K}{}_{MN}\,B_{\mu\nu}^{MN} &\equiv&
-\ft12\,\Theta^{K}{}^{\alpha}\,B_{\mu\nu\,\alpha}\;,
\qquad {\rm with}\quad
B_{\mu\nu\,\alpha}=(t_{\alpha})_{MN}\,B_{\mu\nu}^{MN}
\;.
\eea
They thus can be labeled by indices in the adjoint representation of 
the global symmetry group~${\rm G}$.
In other words, out of the two-forms $B_{\mu\nu}^{MN}$
in the symmetric tensor product\footnote{The fact
that the adjoint representation of ${\rm G}$ appears in the
symmetric tensor product of the vector field representation
is just another way of expressing the fact that in $D=4$
dimensions the group ${\rm G}$ is embedded
into ${\rm Sp}(m,m)$ as discussed
in section~\ref{subsec:even}.
}
\bea
({\cal R}_{\rm v} \otimes {\cal R}_{\rm v})_{\rm sym} &=&
{\cal R}_{\rm adj}\oplus~\dots
\;,
\label{sym4D}
\eea
only those transforming in the adjoint representation
${\cal R}_{\rm adj}$ are involved in the gauging.
This is precisely in accordance with the fact that two-forms in 
four dimensions are dual to the 
scalar field isometries as a consequence of the 
on-shell duality (\ref{duality-scalars})
and thus transform in the adjoint representation of ${\rm G}$. 
The argument shows the need for the linear representation
constraint~(\ref{linear4D}) in $D=4$ dimensions from 
purely bosonic considerations, ensuring that the gauge algebra can be rendered
consistent by adding precisely the two-forms $B_\alpha$ that we have
at our disposal.
On the other hand, an embedding tensor
satisfying $X_{(MNL)}\not=0$ would require the coupling of more two-forms
(in the full symmetric tensor product (\ref{sym4D}))
for consistency of the tensor gauge algebra,
in contradiction with the given field content.

To summarize, we have seen in this section that a general gauging not
only requires covariant derivatives~(\ref{covariant})
but also St\"uckelberg-type couplings between vector 
fields and two-form tensors in order to define the
covariant field strengths~(\ref{covH}).
Consistency of the tensor gauge algebra
poses linear representation constraints~(\ref{linearconstraint}) 
on the embedding tensor,
i.e.\ restricts the possible gaugings.
In fact, the presence of two-form tensor fields in the effective actions precisely 
fits with what is observed in explicit flux compactifications,
e.g.\ the massive two-form field appearing in particular flux compactifications
on Calabi-Yau manifolds~\cite{Louis:2002ny}.

Let us close this section with a few remarks 
on the vector/tensor gauge transformations~(\ref{gaugeAB})
\begin{itemize}
\item
Strictly speaking, the variation of $B_{\mu\nu}^{MN}$ as it stands 
in (\ref{gaugeAB}) is only exact under
projection with the tensor $Z^{P}{}_{MN}$\, --- 
this is for instance of importance when 
verifying closure of the gauge algebra~(\ref{algebra}) below.

\item
The gauge transformations~(\ref{gaugeAB}) close into the algebra
\bea
  {}[\delta(\Lambda_1),\delta(\Lambda_2)] &=& \delta(\Lambda) +
  \delta(\Xi) \;, 
  \label{algebra}
\\[1ex]
\mbox{with}
&&{}
  \Lambda{}^M = g\,X_{[NP]}{}^M \,
  \Lambda_1^N\Lambda_2^P\,, \nonumber\\[.5ex]
&&{}  \Xi_{\mu}^{MN} =  \Lambda_1^{(M}
  D_\mu\Lambda_2^{N)} -
  \Lambda_2^{(M} D_\mu \Lambda_1^{N)} 
  \;,
\nonumber
\eea
showing once more the need of introducing the extra
St\"uckelberg shift $\Xi_{\mu}^{MN}$
on the vector fields in~(\ref{gaugeAB})
in order to close the gauge algebra.
In the presence of higher-rank tensor fields,
the r.h.s.\ of (\ref{algebra}) in general contains also 
the corresponding higher-rank tensor gauge transformations,
whose action on the vector fields is trivial.

\item
Naively, one might have expected 
the standard homogeneous
transformation behavior
$\delta B^{MN}=-2g\Lambda^KX_{KL}{}^{(M} B^{N)L}$ 
of the two-forms under gauge transformations,
rather than the covariant $\Lambda^{(M}{\cal H}^{N)}$ 
term in~(\ref{gaugeAB}). However, the latter contains a contribution
$-2g\Lambda^{{(M}}{Z}^{{N)}}{}_{PQ}\,B^{PQ}$
which comes close to the above homogenous term.
The discrepancy precisely vanishes under
projection with $Z^R{}_{MN}$ as a consequence
of the identity
\bea
Z^R{}_{MP}X_{SQ}{}^{M} 
+Z^R{}_{MQ}X_{SP}{}^{M} -
2Z^R{}_{MS}{Z}^{{M}}{}_{PQ}
&=&0
\;,
\eea
which in turn follows as a consequence of the 
quadratic constraint~(\ref{closure2}).
Under projection thus one recovers from~(\ref{gaugeAB})
the standard homogenous transformation behavior.
Most importantly, the full field strengths
${\cal H}^{M}$ transform covariantly under the 
gauge transformations~(\ref{gaugeAB}).

\item
The full covariant field strength~(\ref{covH}) no longer 
satisfies the standard Bianchi identies, but rather
its deformed version
\bea
D^{\vphantom{M}}_{[\mu}{\cal H}^M_{\nu\rho]} &=&
\ft13\,gZ^M{}_{PQ}\,{\cal H}^{PQ}_{\mu\nu\rho}
\;,
\label{defBianchi}
\eea
where ${\cal H}^{PQ}_{\mu\nu\rho}$
denotes the covariant field strength of the two-forms.

\item
The combined transformation
\bea
\Xi_\mu^{{MN}}  = D_\mu\,\xi^{MN} \;,
\qquad
\Lambda^{M} = g Z^{M}{}_{{PQ}}\,\xi^{PQ}
\;,
\eea
has no effect on the gauge fields $A_\mu^M$, $B_{\mu\nu}^{MN}$
(the latter again under projection with $Z^{P}{}_{MN}$).
This is the proper 
non-abelian generalization of
the standard tensor gauge redundancy.

\item
The structure we have presented here for vector fields
and two-forms
extends to the full hierarchy
of higher-rank $p$-forms.
In particular, the full set of vector/tensor gauge transformations 
takes the form (schematically)
\bea
\delta {\cal V} &=& g \Theta \,\Lambda {\cal V}
\;,
\nonumber\\
\delta A_{\mu} &=& D_{\mu}\Lambda ~- g \Theta \,\Xi_{\mu}  
\;,
\nonumber\\
\delta B_{\mu\nu} &=& 2D_{[\mu}\Xi_{\nu]} ~+~ \dots ~- g \Theta \,\Phi_{\mu\nu}  
\;,
\nonumber\\
\delta C_{\mu\nu\rho} &=& 3D_{[\mu}\Phi_{\nu\rho]} ~+~ \dots ~- g \Theta \,\Sigma_{{\mu\nu\rho}}  
\;,
\nonumber\\[.3ex]
\mbox{etc.}\;,
&&
\label{hierarchy}
\eea
where we have ommitted all ${\rm G}$-indices.
This shows the non-trivial entanglement between $p$-forms and $(p-1)$-forms
via St\"uckelberg terms induced by a generic gauging (i.e.\ a generic tensor $\Theta$).
We have schematically denoted all the intertwining tensors by $\Theta$
as they are uniquely defined in terms of the embedding tensor,
while their precise index structure takes care of the different representations
in which the $p$-forms transform
(with $Z^K{}_{MN}$ from (\ref{defZ}) as the lowest explicit intertwining tensor).
The representation content of the embedding tensor is
determined from the linear constraint (\ref{linearconstraint}) which we have seen
to follow from consistency of the tensor gauge algebra on the lowest-rank tensor fields.
Remarkably, a closer analysis of the higher-rank tensor gauge 
transformations~(\ref{hierarchy}) then allows to determine the representation 
content of all higher-rank $p$-forms of the theory, 
see~\cite{deWit:2005hv,deWit:2008ta}.
For the maximal supergravities, with the embedding tensor
given in table~\ref{table:embedding}, this reproduces the entire
field content of these theories, including the non-propagating $(D-1)$ and $D$-forms.
In particular, it gives agreement with the predictions obtained
from analyzing the branchings of the infinite-dimensional representations
of the underlying very extended Kac-Moody algebras,
\cite{Riccioni:2007au,Bergshoeff:2007qi,Bergshoeff:2007vb,Riccioni:2007ni}.

\end{itemize}

The task in the following will be to put all these structures 
on the level of the Lagrangian.

%%%%%%%%%%%%%%%%%%%%%%%%%%%%%%%%%%%%%%%
\subsection{The Lagrangian}
\label{subsec:Lagrangian}
%%%%%%%%%%%%%%%%%%%%%%%%%%%%%%%%%%%%%%%

In this section we will describe how
to obtain the Lagrangian that is compatible with the 
new local symmetry~(\ref{gaugeAB})
as a deformation of the Lagrangian of 
the ungauged theory.
Given the algebraic framework we have set up
in the last two sections, the first step obviously
consists of covariantizing all derivatives
according to (\ref{covariant})
and to replace the abelian field strengths
by the full covariant ones (\ref{covH})
and their analogues for the higher-rank $p$-forms.
It is slightly more tedious but straightforward to
also covariantize the topological terms present in
the ungauged theory.

We have seen in the last section, that the deformation
leads to an entanglement of the $p$-forms and the $(p+1)$-forms
via the corresponding field strengths. As a consequence,
the covariantized Lagrangian will carry forms of higher degree
than the ungauged one. E.g.\ in $D=4, 5$ dimensions,
the gauged theory generically carries 2-forms, in $D=6, 7$
dimensions the gaugings carry 3-forms, etc., see table~\ref{table:pforms-gauged}
(cf.~in contrast table~\ref{table:pforms} for the ungauged theories).
Moreover, since the construction of the deformation is 
manifestly ${\rm G}$-covariant, the gauged theories in even dimensions
generically carry the full ${\rm G}$-representation of forms
rather that only the electric half.

\begin{table}[tb]
\centering
\begin{tabular}{r|c||c|c|c|c|}
$D$ &${\rm G}$&  $1$&$2$&$3$ &$4$ \\   \hline
9   & ${\rm GL}(2)$   
& ${\bf 1}^{-4} + {\bf 2}^{+3}$ & ${\bf 2}^{-1}$ & ${\bf 1}^{+2}$& ${\bf 1}^{-2}$\\
8   & ${\rm SL}(2)\! \times \!{\rm SL}(3)$  & 
 $({\bf 2},{\bf 3}')$&  $({\bf 1},{\bf 3})$&  $({{\bf 2}},{\bf 1})$ & $({\bf 1},{\bf 3}')$
 \\
7   & ${\rm SL}(5)$ & ${\bf 10}'$  & ${\bf 5}$ &  ${\bf 5}'$& \\
6  & ${\rm SO}(5,5)$  & ${\bf 16}_c$ & ${\bf 10}$ &${\bf 16}_s$&\\
5   & ${\rm E}_{6(6)}$ & ${\bf 27}'$ & ${\bf 27}$& &\\
4   & ${\rm E}_{7(7)}$ & ${\bf 56}$ & ${\bf 133}$ && \\
3   & ${\rm E}_{8(8)}$ &${\bf 248}$ & &&
\\ 
\end{tabular}
\caption{\small
The $p$-forms $(p\ge1)$ entering the Lagrangian of gauged maximal supergravity.
}\label{table:pforms-gauged}
\end{table}

A priori, the presence of these extra fields in the gauged theory
might pose a formidable obstacle to the construction: as
these fields were not present in the ungauged theory, they
do not possess kinetic terms but instead only appear as corrections
to lower-rank field strengths and topological terms upon covariantization.
This might lead to weird if not inconsistent additional field equations.
Instead, somewhat miraculously, it turns out that the various contributions 
from kinetic and topological terms
precisely combine into {\em first-order} field equations for the additional fields.
This reflects the fact that these fields do not constitute additional
degrees of freedom but are the on-shell duals of the fields
of the ungauged theory.
From this perspective, remarkably, this reasoning gives a purely bosonic 
argument for the appearance of the topological terms in the ungauged theory: 
it is their covariantization that renders the field equations
in the gauged theory consistent.\footnote{
A notable exception is the three-dimensional
theory, whose ungauged version does not carry any topological term,
such that a gauge invariant Chern-Simons term for the vector fields
must be added to the gauged theory, in order to produce
sensible field equations.}
In other words, in absence of the standard topological term
(whose presence is usually deduced from supersymmetry)
the bosonic theories would not allow for generic deformations.

Let us discuss as an example the theories in $D=4$ space-time dimensions.
The general gauging is defined by an embedding tensor $\Theta_M{}^\alpha$
giving rise to covariant derivatives~(\ref{covariant}) which in general involve
all vector fields
\bea
D_\mu ~\equiv~
\partial_\mu - gA_\mu^M\,\Theta_M{}^\alpha\,t_\alpha
~=~
\partial_\mu - gA_\mu^\Lambda\,\Theta_\Lambda{}^\alpha\,t_\alpha
- gA_{\mu\Lambda}\,\Theta^\Lambda{}^\alpha\,t_\alpha
\;,
\label{electric-magnetic}
\eea
where according to the discussion of section~\ref{subsec:even}
we have split the $2m$ vector fields into the $m$ electric ones $A_\mu^\Lambda$ and 
their magnetic duals. Only the former ones appear in the ungauged theory.
While at first sight it may seem unnatural to include the magnetic vector fields
in the general connection, this is in fact indispensable in order to achieve
a duality covariant description of flux compactifications.
Recall that upon different compactification from higher dimensions
one usually ends up in different symplectic frames in four dimensions. I.e.\
depending on the higher-dimensional context, the effective
four-dimensional theory might carry different selections of $m$ electric
vector fields among the $2m$ gauge fields.
Only after exchanging some electric versus magnetic vector fields, one may
be able to pin down the equivalence/duality between different compactifications.
This shows that a restriction to electric vector fields in (\ref{electric-magnetic})
might miss certain effective theories which correspond to standard electric 
gaugings in another symplectic frame.
On the other hand, with all components of $\Theta_M{}^\alpha$ present in (\ref{electric-magnetic})
it is straightforward to identify the action of the duality group onto the various
flux parameters in different compactifications. 
We shall see this in more detail in the last section.

The appearance of the $A_{\mu\,\Lambda}$
in the covariant derivatives~(\ref{electric-magnetic})
could lead to the problems discussed above:
as new fields they seem to appear in the role of
Lagrange multipliers that would imply some devastating field equations.
However, the above discussed mechanism comes to the rescue:
gauge invariance of the Lagrangian~(\ref{Leven}), also requires
the introduction of additional topological terms of the form~\cite{deWit:2005ub}
\bea
{\cal L}_{\rm top} &\propto&
\varepsilon^{\mu\nu\sigma\tau}\,\Big(
g\Theta^{\Lambda\,\alpha}\,\partial_\mu A_{\nu\,\Lambda}\,
B_{\sigma\tau\,\alpha}
+\ft18
g^2\,
\Theta^{\Lambda\,\alpha}\Theta_{\Lambda}{}^{\beta}
\,
B_{\mu\nu\,\alpha}
B_{\sigma\tau\,\beta}
+ \dots
\Big)
\;.
\eea
Together, it follows from 
variation of $B_{\mu\,\alpha}$ that
\bea
g\,\Theta^{\Lambda\,\alpha}\,
\Big({\cal H}_{\mu\nu\,\Lambda}
+e\varepsilon_{\mu\nu\sigma\tau}\,
\frac{\delta{\cal L}_{\rm kin}}{\delta {\cal H}_{\sigma\tau}{}^{\Lambda}}
\Big)&=& 0
\;,
\label{eomdual}
\eea
which precisely reproduces the covariant version of
the duality equation~(\ref{defGeven}).
Likewise, variation w.r.t.\ the magnetic vector fields induces
the duality equation~(\ref{duality-scalars}) between scalars and two-forms.
Note that for the gauged theory, the duality equations arise
as true field equations, however projected with the 
matrix $\Theta^{\Lambda\,\alpha}$.
In particular, in the limit $g\rightarrow0$, all dual fields disappear from the 
Lagrangian and equation~(\ref{eomdual}) consistently decouples.
This is different from the {\em democratic} formulation of supergravities, 
in which the dual fields are introduced already in the ungauged action
and the duality relations~(\ref{defGeven}), etc. must be supplied by hand.

Summarizing, we have succeeded in finding a deformation of
the original ungauged Lagrangian of supergravity that is 
compatible with the algebraic structures induced by the 
new local gauge group and encoded in the embedding tensor $\Theta_M{}^\alpha$
as presented in the last sections.
Details of the construction may differ in the various space-time dimensions 
and can be found in the literature for several 
examples, see e.g.~\cite{Nicolai:2001sv,deWit:2004nw,deWit:2005ub,
Samtleben:2005bp,Schon:2006kz,deRoo:2006ms,Derendinger:2007xp,
Samtleben:2007an,deVroome:2007zd,
deWit:2007mt,Bergshoeff:2007ef}.\footnote{In dimensions $D=8$ and $D=9$,
gauged supergravities have been classified and constructed independently, 
based on the different
compactification manifolds~\cite{Bergshoeff:2002nv,Bergshoeff:2003ri}.
While the nine-dimensional case is in exact agreement with the form
of the embedding tensor of table~\ref{table:embedding}, 
the compactifications to eight dimensions seem to reproduce only part
of the possible components of the corresponding embedding tensor.
}
The fact that the deformation has been described in a manifestly
${\rm G}$-covariant way has another appealing consequence:
in even dimensions --- where ${\rm G}$ is realized only on-shell --- 
this construction can accommodate gaugings
of subgroups ${\rm G}_0$ of ${\rm G}$ that are 
not among the off-shell symmetries of the ungauged Lagrangian!

Now, that we have constructed a gauge invariant Lagrangian,
we may take the next step and check if the deformation is further
compatible with local supersymmetry.
As it stands, the deformed Lagrangian is no longer 
invariant under supersymmetry due to the extra contributions
that arise from variation of the vector fields in the covariant derivatives,
and from the deformation of Bianchi identities (\ref{defBianchi}), etc.
in the gauged theories.
Supersymmetry can be restored
applying the standard Noether procedure~\cite{deWit:1982ig}.
In linear order of the deformation parameters $\Theta$,
the unwanted contributions can be cancelled by
introducing particular fermionic mass terms
of the type (schematically)
\bea
{\cal L}_{{\rm ferm-mass}}
&=&
g
\,\Big(
\overline{\psi}{}^i\,A_{ij}\,\psi^j+
\overline{\chi}{}^A\,B_{Ai}\,\psi^i+
\overline{\chi}{}^A\,C_{AB}\,\chi^B
\Big)~+~ {\rm h.c.}
\;,
\label{ferm-mass}
\eea
where by $\psi^i$ and $\chi^A$ we denote gravitinos and
spin-$1/2$ fermions, respectively, with the indices $i$ and $A$
labeling the respective ${\rm K}$-representations,
and where we have suppressed all space-time indices and 
$\gamma$-matrices.\footnote{Our treatment of fermions
will remain somewhat schematic in this section as we are trying to
give a discussion for arbitrary theories,
whereas e.g.\ their symmetry and hermiticity properties
certainly depend on the number $D$ of space-time dimensions
and $N$ of supersymmetries, see e.g.~\cite{Strathdee:1986jr,Tanii:1998px,deWit:2002vz} 
for a discussion of spinor fields in various dimensions.
}
The tensors $A_{ij}$, $B_{Ai}$, and $C_{AB}$ may depend on the
scalar fields, and inherit their symmetry properties from their
precise appearance in~(\ref{ferm-mass}). Under 
the action~(\ref{symmGK}) of ${\rm K}$ 
they should transform
in the proper representations such that~(\ref{ferm-mass})
is ${\rm K}$-invariant.
Together with the fact that these tensors must be defined in terms
of the embedding tensor $\Theta_M{}^\alpha$ that encodes the deformation,
their transformation properties entirely fix the form of these tensors.
Specifically, they are constructed from the so-called $T$-tensor defined by
\bea
T_{\underline{N}}{}^{\underline{\beta}} &\equiv&
\Theta_M{}^\alpha\,{\cal V}^M{}_{\underline{N}}\,
{\cal V}_\alpha{}^{\underline{\beta}}
\;,
\eea
as the embedding tensor dressed with the scalar group matrix 
${\cal V}$ evaluated in the fundamental and the adjoint representation
of ${\rm G}$, respectively. 
This object has first appeared in the ${\rm SO}(8)$ gauging in $D=4$
dimensions~\cite{deWit:1982ig}.
In contrast to the constant $\Theta_M{}^\alpha$, the $T$-tensor
depends non-trivially on the scalar fields.
As it is obtained from the embedding tensor by a finite 
${\rm G}$ transformation, it lives in the 
same ${\rm G}$-representation as $\Theta$,
i.e.\ it inherits from $\Theta$ the linear constraint~(\ref{linearconstraint})
\bea
\mathbb{P}\:T &=& 0 \;,
\eea
which now holds for any value of the scalar fields on which 
$T$ depends.
Under ${\rm K}$ this tensor contains various irreducible parts,
obtained by decomposing the ${\rm G}$-representation 
of~$\Theta$ (cf. table~\ref{table:embedding}) under the compact subgroup ${\rm K}$.
These ${\rm K}$-irreducible parts can precisely be identified with 
the fermionic mass tensors in~(\ref{ferm-mass}).

For example in $D=4$, ${\cal N}=8$, under ${\rm K}={\rm SU}(8)$ the embedding tensor
breaks into
\bea
\Theta_M{}^\alpha
&\longrightarrow& ~T_{\underline{M}}{}^{\underline{\alpha}} 
~\rightarrow~
(\,A^{ij} ,\;\; A_{ij} ,\;\; B^{Ai} ,\;\; B_{Ai}\;)
\;,
\eea
according to the decomposition
${\bf 912} \rightarrow
{\bf 36} + \overline{{\bf 36}}
+{\bf 420} + \overline{{\bf 420}}\,$,
from which the fermionic mass tensors are built.
For the explicit formulas of the tensors 
$(A^{ij}, A_{ij}, B^{Ai}, B_{Ai})$
in terms of a general $\Theta$, we refer 
to~\cite{deWit:2007mt}.

Turning the argument around, this shows the 
origin of the linear representation
constraint from supersymmetry.
The supersymmetry-violating terms in the Lagrangian 
which are proportional to the embedding tensor, e.g.\ as (schematically)
\bea
F^M_{\mu\nu} \, \Theta_M{}^\alpha\,
{\cal V}_\alpha{}^{\underline{\beta}} \; (\bar\epsilon\,\psi)_{\underline{\beta}}
&=&
(F^M_{\mu\nu} \,{\cal V}_M{}^{\underline{N}})
\,T_{\underline{N}}{}^{\underline{\beta}} \;
 (\bar\epsilon\,\psi)_{\underline{\beta}}
 \;.
 \label{unwanted}
\eea
can be cancelled by the variation of the additional fermionic mass
terms~(\ref{ferm-mass}) if and only if the tensor 
$T_{\underline{N}}{}^{\underline{\beta}}$\,
can be built from the representations of proper fermionic mass tensors.
In $D=4$, ${\cal N}=8$, the possible fermionic mass tensors fall into 
${\rm SU}(8)$ representations
\bea
(\bar\psi \psi): &&({\bf 8} \otimes {\bf 8})_{\rm sym} = {\bf 36}\;,
\non
(\bar\psi \chi):&& ({\bf 8} \otimes \overline{\bf 56})~ = 
{\bf 28}+\overline{\bf 420} \;,
\non
(\bar\chi \chi): &&(\overline{\bf 56} \otimes \overline{\bf 56})_{\rm sym} = 
{\bf 420}+\overline{\bf 1176}
\;,
\label{fermionicmass}
\eea
and their hermitean conjugates.
Comparing this to the possible representation content 
of a generic embedding tensor~$\Theta_M{}^\alpha$ from~(\ref{912})
\bea
{\bf 56} &\rightarrow& {\bf 28}~+ {\rm\bf h.c.}\;,
\non
{\bf 912} &\rightarrow& {\bf 36} + {\bf 420} ~+ {\rm\bf h.c.}\;,
\non
{\bf 6480} &\rightarrow&  {\bf 28}+ {\bf 420} +
 {\bf 1280} + {\bf 1512} ~+ {\rm\bf h.c.}
 \;,
\eea
shows that an embedding tensor 
$\Theta$ in the ${\bf 6480}$ gives rise to 
terms of the type~(\ref{unwanted})
with a part of $T_{\underline{N}}{}^{\underline{\beta}}$
in the ${\bf 1280}+{\bf 1512}$
which cannot be cancelled by fermionic mass 
terms~(\ref{fermionicmass}).
This is the underlying reason why supersymmetry requires 
the linear constraint~(\ref{linearconstraint})
and forbids a $\Theta$ in the ${\bf 6480}$.
Similarly, a $\Theta$ in the ${\bf 56}$ is ruled out by supersymmetry:
although a ${\bf 28}$ appears in $(\bar\psi \chi)$,
a closer check shows that its absence in $(\bar\psi \psi)$
forbids this representation in the embedding tensor.
Moreover, (\ref{fermionicmass}) shows that in this theory
the mass tensor $C_{AB}$ of the spin-$1/2$ fermions is in fact 
obtained from the mixed mass tensor $B^{Ai}$ --- as there is 
only a single ${\bf 420}$ contribution within the ${\bf 912}$.
Indeed, this was first discovered in the ${\rm SO}(8)$ gauging
of~\cite{deWit:1982ig}.

Let us recall that in the previous sections we have found
linear representation constraints on the embedding tensor
from purely bosonic considerations --- consistency of the 
deformed $p$-form tensor hierarchy.
It is remarkable and somewhat surprising
that supersymmetry appears to impose 
precisely the same linear constraint on the possible 
deformations such that no further restriction descends 
from compatibility with supersymmetry.\footnote{
Again, a notable exception is the three-dimensional theory,
where supersymmetry imposes linear constraints 
on the embedding tensor that do not already follow from 
consistency of the bosonic Lagrangian~\cite{deWit:2003ja}.}

If the embedding tensor satisfies the linear representation
constraint, the additional fermionic mass terms~(\ref{ferm-mass}) are 
precisely sufficient to
cancel all supersymmetry-violating terms in linear order of $\Theta$ if
simultaneously the fermionic supersymmetry transformations are modified
according to (schematically)
\bea
\delta \psi^i &=& \delta_0 \psi^i - gA^{ij}\,\epsilon_j
\;,
\qquad
\delta \chi^A ~=~ \delta_0 \chi^A - gB^{Ai}\,\epsilon_i
\;,
\label{fermion-shifts}
\eea
where $\delta_0$ denotes the (properly covariantized)
supersymmetry transformations of the ungauged theory.
The reason for the additional fermion-shifts in (\ref{fermion-shifts})
is to cancel the $D_\mu\epsilon$ contributions descending from~(\ref{ferm-mass}).

Finally, supersymmetry in second order $g^2$ of the deformation requires
the addition of a scalar potential, which is schematically of the form
\bea
{\cal L}_{\rm pot} &=&
-eV ~=~ -e g^2\,\Big( B^{Ai}B_{Ai} -A^{ij}A_{ij} \Big)\;,
\label{potential-general}
\eea
in terms of the fermionic mass tensors,
in order to cancel the $g^2$ contributions descending from the action
of (\ref{fermion-shifts}) on~(\ref{ferm-mass}). 
It is characteristic for supergravity theories that --- in contrast 
to globally supersymmetric theories --- the scalar potential
is in general not positive definite, but may in particular support 
dS and AdS vacua.
For particular gaugings, i.e.\ particular choices of $\Theta$,
despite its appearance of (\ref{potential-general}), 
the potential may still be positive definite.

Consistent cancellation of all supersymmetry variations in order $g^2$
typically requires a number of nontrivial algebraic identities to be
satisfied by the fermionic mass tensors $A_{ij}$, $B_{Ai}$, and 
$C_{AB}$. In particular, one needs the 
traceless condition
\bea
g^2\,\Big( B^{Ai}B_{Aj} -A^{ik}A_{jk} \Big)  &\equiv&
\ft1{N}\,\delta^i_j \,V
\;,
\label{susyWard}
\eea
with $N$ the number of supercharges and
the scalar potential $V$ from~(\ref{potential-general})
--- often referred to as a supersymmetric Ward identity.
As this is a condition which is bilinear in the embedding tensor,
the only way it can be satisfied without imposing further constraints
on the gauging is as a consequence of the quadratic 
constraint~(\ref{quadraticconstraint}). Indeed, in all dimensions,
(\ref{susyWard}) and analogous relations can be derived 
from~(\ref{quadraticconstraint})
upon dressing the latter with the scalar matrix  ${\cal V}$
and breaking it into its ${\rm K}$-irreducible parts.

It is sometimes convenient to express the scalar potential
directly in terms of the embedding tensor rather than going through the
process of computing the fermionic mass tensors. E.g.\ for the 
maximal ${\cal N}=8$ theory in $D=4$ dimensions, the potential takes the
equivalent form~\cite{deWit:2007mt}
\bea
V &=& 
g^2\,\Big(
X_{MN}{}^{R}\,X_{PQ}{}^{S}\,{\cal M}^{MP}{\cal M}^{NQ}{\cal M}_{RS} 
+ 7\,X_{MN}{}^{Q}\,X_{PQ}{}^{N}\,{\cal M}^{MP} 
\Big)\;,
\label{potentialN8}
\eea
with $X_{MN}{}^K$ defined in (\ref{closure}) as a function of the embedding tensor,
and the positive definite scalar matrices ${\cal M}_{MN}$ defined in (\ref{defM}).
This provides a universal and very compact form for the scalar potential
obtained in generic flux compactifications. Depending on the particular fluxes 
present in the compactification, different blocks of the embedding tensor will
be non-vanishing and shape the dependence of $V$ on the scalar fields
contained in ${\cal M}_{MN}$. We will come back to this in section~\ref{sec:flux}.

In general, the scalar potential can always be cast into the form
\bea
V&=&g^2\, {\rm V}^{MN}{}_{\alpha\beta}{}\,\Theta_M{}^\alpha\,\Theta_N{}^\beta
\;,
\eea
in terms of the embedding tensor and a scalar dependent matrix 
$ {\rm V}^{MN}{}_{\alpha\beta}$ which e.g.\ 
for the maximal $N=8$ theory can be extracted from 
(\ref{potentialN8}).
Interestingly, this matrix shows up in 
the analogue of the duality relations~(\ref{duality})
for the $(D-1)$ forms of the theory.
These non-propagating forms, whose field content can e.g.\ be 
deduced from the supersymmetry algebra of the 
ungauged theory \cite{Bergshoeff:2005ac,Riccioni:2007ni,deWit:2008ta}
in general transform in the representation dual to the embedding tensor,
i.e.~carry indices of the type $C^M{}_\alpha$. 
Whereas these forms are usually set to zero in the ungauged theory,
their ${\rm G}$-covariant equations of motion can be integrated 
to~\cite{deWit:2008ta}
\bea
\partial_{[\mu_1} (C^M{}_\alpha)_{\mu_2\dots\mu_D]}
~+~ \cdots 
&=& 
 e\,\varepsilon_{\mu_1 \dots \mu_D}\,{\rm V}^{MN}{}_{\alpha\beta}{}\,\vartheta_N{}^\beta
 \;,
\eea
with integration constants $\vartheta_N{}^\beta$ and the dots representing possible 
Chern-Simons contributions to the field strength.
Non-vanishing integration constants in this equation precisely 
induce the gauged theory
with the identification $\vartheta_N{}^\beta\equiv\Theta_N{}^\beta$.

%%%%%%%%%%%%%%%%%%%%%%%%%%%%%%%%%%%%%%%
%%%%%%%%%%%%%%%%%%%%%%%%%%%%%%%%%%%%%%%

\section{Flux Compactifications --- Examples}
\label{sec:flux}

%%%%%%%%%%%%%%%%%%%%%%%%%%%%%%%%%%%%%%%
%%%%%%%%%%%%%%%%%%%%%%%%%%%%%%%%%%%%%%%

In this final section we will work out a few explicit examples 
of four-dimensional 
gauged supergravities associated to particular flux compactifications
along the lines discussed at the end of section~\ref{subsec:embedding}.
The maximal and half-maximal supergravities are in particular relevant
for flux compactifications on tori (and their orientifolds)
and we will mainly consider the torus
compactifications from M-theory and the IIA/IIB theories.

The simplest flux compactifications refer to compactifications with 
non-trivial values
\bea
\int_{\Sigma}\,{\cal F}^{(p)} &=& {\cal C}_\Sigma
\;.
\label{flux}
\eea
of $p$-form field strengths ${\cal F}$ along non-trivial cycles $\Sigma$
of the internal manifold.
The constants ${\cal C}_\Sigma$ can be considered as deformation parameters
and as such be identified within the components of the embedding tensor $\Theta_M{}^\alpha$ introduced above.
In the following, we will mainly consider compactifications on tori $T^n$,
where the non-trivial cycles are products of circles and thus labeled by
indices along the directions of the torus.

%%%%%%%%%%%%%%%%%%%%%%%%%%%%%%%%%%%%%%%
\subsection{Higher-dimensional origin of symmetries}
\label{subsec:origin}
%%%%%%%%%%%%%%%%%%%%%%%%%%%%%%%%%%%%%%%

A crucial role in the covariant construction of gaugings was
played by the underlying global symmetry groups ${\rm G}$ of the 
ungauged theories, given in table~\ref{table:groups}.
In order to work out the gaugings which correspond to
the effective theories of particular flux compactifications,
it will thus be important to first understand the higher-dimensional
origin of these symmetry groups.
Recall, that the ungauged maximal and
the half-maximal theories arise from
reduction of eleven- and ten-dimensional supergravity, respectively, on an $N$-torus,
with the global symmetry groups given by 
exceptional and the orthogonal 
series ${\rm E}_{N(N)}$ and ${\rm SO}(N,N)$, respectively.
In both cases, the maximal ${\rm GL}(N)$ subgroups have a
relatively simple higher-dimensional interpretation related to the 
geometry of the $N$-torus,
while the remaining part of the groups is related to higher-dimensional tensor fields.

Let us first consider the reduction of pure gravity from $(D+N)$ dimensions
down to $D$ dimensions.
With coordinates splitting according to
$x^M\rightarrow (x^\mu, y^m)$,\, $\mu=0, \dots, D-1$;\,$m=1, \dots N$,
and similarly for the flat indices $A\rightarrow (\alpha, a)$,
the reduction ansatz for the vielbein on an $N$-torus is
given by
\bea
E_M{}^A &=&
\left(
\begin{array}{cc}
{\rm e}^{\kappa\phi}\,e_\mu{}^\alpha & {\rm e}^{\phi/N}\,V_m{}^a\,B_\mu^m \\
0 & {\rm e}^{\phi/N}\,V_m{}^a
\end{array}
\right)
\;,
\label{vielbeinreduced}
\eea
with all components depending only on the coordinates $x^\mu$.
The matrix $V_m{}^a$ is normalized by ${\rm det}\,V=1$, and $\kappa=\frac{1}{2-D}$ 
is chosen such that the lower dimensional
action appears again in the Einstein frame.
The $D$-dimensionsal theory thus carries a vielbein, $N$ vector fields
and $N^2$ scalar fields. 
The ansatz~(\ref{vielbeinreduced}) preserves an ${\rm SO}(1,D\!-\!1)\times {\rm SO}(N)$
subgroup of the original Lorentz group. The second factor can be used to remove
$\frac12N(N-1)$ of the components in $V_m{}^a$ by virtue of
\bea
\delta V_m{}^a &=& V_m{}^b \,h^a{}_b\;,\qquad
h\in \mathfrak{so}(N)\;,
\label{Lorentz}
\eea
leaving $\frac12N(N+1)$ physical scalars in the reduced theory.

The diffeomorphism symmetries $\xi^M$ 
of the $(D+N)$-dimensional theory 
induce different symmetries in the reduced theory.
While diffeomorphisms $\xi^\mu(x)$ induce $D$-dimensional diffeomorphisms,
it is easy to check that the diffeomorphisms $\xi^m(x)$ along the compactified 
directions induce abelian gauge
transformations for the Kaluza-Klein vector fields
\bea
\delta B_\mu^m &=& \partial_\mu \xi^m\;.
\eea
Moreover, diffeomorphisms linear in the $N$ compactified coordinates,
$\xi^m=-\Lambda^m{}_n\,y^n$, with a traceless matrix $\Lambda$,
induce a global ${\rm SL}(N)$ symmetry acting as
\bea
\delta_\Lambda V_m{}^a &=& \Lambda^n{}_m\,V_n{}^a \;,\qquad
\delta_\Lambda B^m_\mu ~=~ -\Lambda^m{}_n\,B^n_\mu
\;,
\label{actionSL}
\eea
on the components of (\ref{vielbeinreduced}).
Diffeomorphisms corresponding to constant rescaling of the $N$-torus, 
$\xi^m=\lambda\,y^m$, are slightly more delicate. 
A priori they induce an action (\ref{actionSL})
with diagonal matrix $\Lambda$. But as they also induce an
action on the $D$-dimensional vielbein~$e_\mu{}^\alpha$, 
they do not constitute an off-shell symmetry in $D$ dimensions. 
However, combined with a proper rescaling of the 
$(D\!+\!N)$-dimensional vielbein~(\ref{vielbeinreduced})
they result in an off-shell ${\rm GL}(1)$ symmetry 
\bea
\label{torus-scale}
\delta_\lambda \phi  &=&  \lambda\,N\,(D-2)\;,\qquad
\delta_\lambda B^m_\mu        ~=~ -\lambda\,(D-2+N) \, B^m_\mu \;,
\label{actionGL}
\eea
of the $D$-dimensional theory, that leaves $e_\mu{}^\alpha$
invariant.
Comparing (\ref{Lorentz}), (\ref{actionSL}), and (\ref{actionGL}) to 
the transformations (\ref{symmGK}), (\ref{actionA}) 
above, we identify the scalar fields as described by an 
${\rm GL}(N)/{\rm SO}(N)$ coset space $\sigma$-model 
with the $N$ vector fields 
transforming in the fundamental representation of the global symmetry ${\rm GL}(N)$. 
The resulting Lagrangian takes the form (\ref{LPP}), (\ref{Lkin}),
discussed in sections~\ref{subsec:scalar} and \ref{subsec:vector} above,
with the matrix ${\cal M}_{mn} \equiv {\rm e}^{2(D-2+N)/(D-2) \,\phi}\,V_m{}^aV_n{}^b\,\delta_{ab}$. 

This completes the structure of pure gravity reduced on
an $N$-torus. The reductions of extended supergravities typically
exhibit larger global symmetry groups into which the ${\rm GL}(N)$
is embedded as a subgroup. The enhancement of the symmetry group 
is related to the presence of additional $p$-form fields in the higher-dimensional
theory. 
The reduction ansatz for these forms is straightforward
\bea
A_{M_1 \dots M_p} &\longrightarrow&
(A_{\mu_1 \dots \mu_p},
A_{m_1\mu_2 \dots \mu_p},
A_{m_1m_2\mu_3 \dots \mu_p},
\dots,
A_{m_1 \dots m_p} )
\;,
\label{reduceforms}
\eea
in terms of $D$-dimensional $p$-forms, $(p-1)$-forms, $(p-2)$-forms, etc.
The transformation behavior of these fields under the 
${\rm SL}(N)$ from (\ref{actionSL})
follows from their index structure in the internal indices $m_1$, $m_2$, \dots,
while for their scaling under the ${\rm GL}(1)$ of (\ref{actionGL})
one obtains (see e.g.~\cite{deWit:2002vt})
\bea
\delta_\lambda A_{m_1\cdots m_k\,\mu_{k+1}\cdots\mu_p}
&=& 
\lambda\,\Big((D-2)\,k+(k-p)\,N\Big)\,
A_{m_1\cdots m_k\,\mu_{k+1}\cdots\mu_p}  \;.
\label{actionGL1}
\eea
In particular, for $N\ge p$ the reduction (\ref{reduceforms})
adds $N\choose p$ scalar fields $A_{m_1 \dots m_p}$
to the $D$-dimensional theory. The higher-dimensional
tensor gauge transformations $\delta A_{M_1 \dots M_p} =
p\,\partial_{[M_1}\,\Xi_{M_2\dots M_p]}$ which are linear in
the compactified coordinates, 
$\Xi_{m_2\dots m_p}=\xi_{m_1 \dots m_p}\, y^{m_1}$, 
induce additional global shift symmetries
\bea
\delta_\xi\, A_{m_1 \dots m_p} &=& 
\xi_{m_1 \dots m_p}
\;,
\label{shift1}
\eea
on these scalar fields.

A final source for scalar fields in the reduced theories
are the $(D-2)$ forms that arise in the reduction (\ref{reduceforms})
(and for $D=3$ also as the Kaluza-Klein 
vector fields in~(\ref{vielbeinreduced})).
As discussed in general in section~\ref{subsec:vector} above, 
in $D$-dimensions these forms can be dualized into scalar fields.
It is important to note that due to their definition 
also these scalar fields $\varphi_a$ obtained
by dualization possess an additional global shift symmetry
\bea
\delta_\zeta\, \varphi_a &=& \zeta_a 
\;.
\label{shift2}
\eea
Together, the symmetries directly inherited from $(D+N)$ dimensions
thus form a non-semisimple group of the type ${\rm GL}(N)\ltimes {\cal N}$,
with nilpotent ${\cal N}$ combining the shifts (\ref{shift1}), (\ref{shift2}).
Typically, these symmetries just
form an upper (Borel) half of the full semi-simple global symmetry group 
${\rm G}$ which may be sketched as
\bea
\resizebox{60mm}{!}{\includegraphics{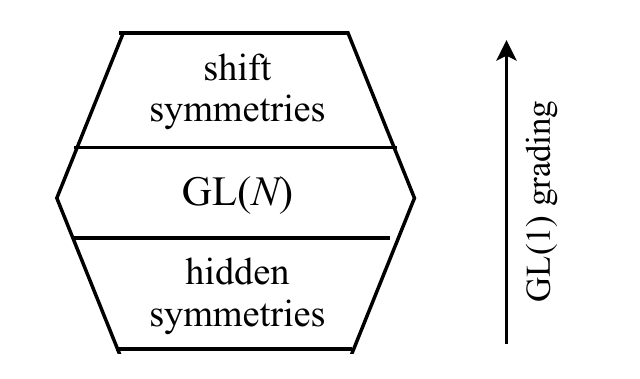}}
\label{decomp}
\eea
In particular, the $D$-dimensional theory typically possesses a number
of (${\rm dim} \,{\cal N}$, to be precise) 
additional symmetries --- often referred to as hidden symmetries ---
that have no obvious higher-dimensional origin, and together with
${\rm GL}(N)\ltimes {\cal N}$ form the semi-simple group ${\rm G}$.
The decomposition~(\ref{decomp}) is along the grading induced
by the ${\rm GL}(1)$ scaling (\ref{actionGL}).
The fact that the number of additional hidden symmetries is precisely enough
in order to form a semi-simple global symmetry group in $D$ dimensions
of course heavily hinges on the field content of the higher-dimensional 
supergravity theory. This is where the underlying supersymmetric structure
that is preserved throughout the reduction comes to play its role.
In the following we shall just make use of this matter of fact for the maximal
and the half-maximal supergravities.
We finally note that the decomposition~(\ref{decomp}) 
naturally selects a Borel subalgebra and thus 
a particular triangular gauge~(\ref{triangular}) for the coset space,
in which the higher-dimensional origin of the $D$-dimensional 
scalar fields becomes most transparent.
Again, we refer to~\cite{Cremmer:1997ct} for a systematic discussion 
of the maximal supergravities in various dimensions and their eleven-dimensional origin.

One of the simplest examples of such a reduction is the 
Kaluza-Klein compactification of simple $D=5$ supergravity on a circle $S^1$.
The bosonic field content of minimal $D=5$ supergravity comprises the metric and
a single vector field which upon reduction~(\ref{vielbeinreduced}), (\ref{reduceforms})
give rise to gravity
coupled to two vectors and two scalar fields.
According to the discussion above, there are two global symmetries
in the four-dimensional theory that are inherited from five dimensions:
the ${\rm GL}(1)$ scaling (\ref{actionGL}) and the shift symmetry (\ref{shift1})
acting on the $A_5$ component of the five-dimensional vector field.
The full global symmetry group in four dimensions is an ${\rm SL}(2)$,
which decomposes as (\ref{decomp}) with each block generated by a single generator.
This precisely corresponds to the example discussed at the end of 
section~\ref{subsec:scalar} with the two scalars parametrizing the coset space
${\rm SL}(2)/{\rm SO}(2)$ and the generators ${\bf h}$, ${\bf e}$, and ${\bf f}$, 
of ${\rm SL}(2)$ corresponding to the scaling, the shift and the hidden symmetry, respectively.

A very different example leading to the same global symmetry group ${\rm SL}(2)$ is provided
by the $S^1$ reduction of Einstein gravity in four dimensions. While the
scaling symmetry is still (\ref{actionGL}), the shift symmetry is of the type (\ref{shift2}) 
and acts on the scalar that is obtained by dualizing the three-dimensional Kaluza-Klein vector. 
It is in this model, that the non-linear action of a hidden symmetry in gravity
(the generator~${\bf f}$ in this example) has first been discovered~\cite{Ehlers:1957}.

To finish this section, let us note that the action of the ${\rm GL}(1)$ scaling symmetry
(\ref{actionGL}), (\ref{actionGL1}) 
is straightforwardly extended onto those components of the higher-dimensional field strength that may serve as flux parameters according to~(\ref{flux}).
In particular, one finds that
\bea
\delta_\lambda {\cal F}_{m_1\cdots m_p}&=&
\lambda\,\Big((D-2)\,p+N\Big)\,
{\cal F}_{m_1\cdots m_p} \;,
\label{actionGL2}
\eea
for those $p$-form field strengths with all indices along the
compactified directions.
This will be relevant in the next sections in order to identify
the proper flux parameters among the components of the embedding tensor.
Similarly, we will in the following consider the theories obtained by 
compactification in the presence of torsion on the internal torus, 
i.e.\ by a deformation of the reduction ansatz~(\ref{vielbeinreduced})
to $E^a = \tilde{E}_m{}^a(x)\,\eta^m(y)$ in the internal part
with the one-forms $\eta^m(y)$ 
satisfying~\cite{Scherk:1979zr,Kaloper:1999yr}
\bea
d \eta^k &=&  {\cal T}^k_{mn}\, \eta^{m}\wedge \eta^{n}
\;.
\label{geoflux}
\eea
with non-vanishing ${\cal T}^k_{mn}$ antisymmetric in the lower indices
--- often referred to as geometric flux. 
Analogously to (\ref{actionGL2}), 
one finds for the ${\rm GL}(1)$ scaling behavior 
of these components
\bea
\delta_\lambda {\cal T}^k_{mn} &=&  
\lambda\,(D-2+N) \,{\cal T}^k_{mn}
\;.
\label{actionGL3}
\eea

%%%%%%%%%%%%%%%%%%%%%%%%%%%%%%%%%%%%%%%
\subsection{M-theory fluxes}
%%%%%%%%%%%%%%%%%%%%%%%%%%%%%%%%%%%%%%%

As a first example, we will study the reduction of eleven-dimensional
supergravity~\cite{Cremmer:1978km} on a seven-torus $T^7$ in the 
presence of fluxes. This example has been analyzed in 
detail e.g.\ in~\cite{DallAgata:2005ff}, 
\cite{Andrianopoli:2005jv,DAuria:2005er,DAuria:2005rv}, 
and~\cite{Hull:2006tp}
The bosonic field content of the eleven-dimensional theory is the metric
and an antisymmetric three-form tensor.
In absence of fluxes, the reduction leads to the maximal
four-dimensional ungauged supergravity~\cite{Cremmer:1979up} which 
has appeared on various occasions in these lectures and carries
28 electric vector fields and 70 scalars described by the coset space
${\rm E}_{7(7)}/{\rm SU}(8)$.

According to the discussion in the previous section, the first step in
understanding the eleven-dimensional origin of the four-dimensional
fields consists of decomposing the four-dimensional global symmetry
group ${\rm E}_{7(7)}$ under the torus ${\rm GL}(7)$.
In terms of ${\rm GL}(7)$ representations, 
the decomposition (\ref{decomp})
takes the form
\bea
{\rm E}_{7(7)} \quad &\longrightarrow&\quad
\begin{array}{c}
{{\bf 7}'}_{\!+4}\\[.2ex]
{\bf 35}_{+2}\\[.2ex] \hline
{\bf 1}_0+{\bf 48}_0\\[.2ex]\hline
{{\bf 35}'}_{-2}\\[.2ex]
{\bf 7}_{-4}
\end{array}
\qquad
\;,
\label{decompE7G7}
\eea
with subscripts indicating 
the charge under ${\rm GL}(1)\subset{\rm GL}(7)$ 
and the ${\bf 1}_0+{\bf 48}_0$ representing the 
adjoint of ${\rm GL}(7)$.
The ${\bf 35}_{+2}$ nilpotent symmetries 
in (\ref{decompE7G7}) correspond to shifts (\ref{shift1}) 
on the 35 scalar fields descending from the eleven-dimensional 
three-form.
The ${{\bf 7}'}_{\!+4}$ shift symmetries act according to (\ref{shift2})
on the scalars that are obtained by dualizing the 7 two-form fields that descend from
the eleven-dimensional three-form.
Their charges can be matched with (\ref{actionGL1}) 
(upon choosing $\lambda=\frac13$).

The 56 vector fields of the four-dimensional theory decompose
according to
\bea
{\bf 56} &\rightarrow&
{\bf 7}'_{-3}+{\bf 21}_{-1}+{\bf 21}'_{+1}+{\bf 7}_{+3}\;,
\label{E7vectors}
\eea
and with the charges from (\ref{actionGL}), (\ref{actionGL1})
one identifies the ${\bf 7}'_{-3}$ and the ${\bf 21}_{-1}$ as the vector fields
descending from the eleven-dimensional metric and the three-form, respectively.
The other fields in (\ref{E7vectors}) represent their magnetic duals in
accordance with the discussion of section~\ref{subsec:even}.

Let us now consider the possible fluxes that can be switched on in this reduction
and their effect in the four-dimensional theory.
The eleven-dimensional three-form tensor field
can acquire a four-form flux~(\ref{flux})
\bea
{\cal F}^{(4)}_{n_1n_2 n_3 n_4} &=& c_{n_1n_2 n_3 n_4}
\;,
\label{F4}
\eea
with indices $n_1$, \dots, $n_4$, running over the seven coordinates of the torus.
As a consequence of the duality~(\ref{duality}),
one may alternatively 
consider a non-vanishing flux for its dual seven-form field strength
\bea
{\cal F}^{(7)}_{n_1\dots n_7} &=& a\,\epsilon_{n_1\dots n_7}
\;.
\label{F7}
\eea
The effective four-dimensional actions
could in principle be determined by explicitly evaluating 
the reduction with the ansatz~(\ref{F4}), (\ref{F7}). 
Rather than going through this
quite lengthy computation, 
we will directly employ the underlying symmetry structure
in order to identify the corresponding theories among the 
general gaugings presented above.
From their index structure and scaling behavior (\ref{actionGL2}),
one reads off that the flux parameters of (\ref{F4}) and (\ref{F7})
transform in the
${\bf 35}'_{+5}$ and ${\bf 1}_{+7}$, respectively, of~${\rm GL}(7)$.
In section~\ref{sec:gauging}, we have established that the general deformation
of the maximal four-dimensional theory is encoded in an embedding tensor
$\Theta_M{}^\alpha$ transforming in the ${\bf 912}$ representation of 
${\rm E}_{7(7)}$.
In order to identify the particular gaugings corresponding to 
the fluxes (\ref{F4}) and (\ref{F7}) we simply have to identify within the ${\bf 912}$ 
these particular ${\rm GL}(7)$ representations.

Breaking the ${\bf 912}$ according to (\ref{decompE7G7}) gives the following set of
representations
\bea
\begin{array}{cl}
{\bf 1}_{+7} & 
\\[.2ex]
{\bf 35}'_{+5} &
\\[.2ex]
{\bf 7}_{+3}+{\bf 140}_{+3} &
\\[.2ex]
{\bf 21}'_{+1}+{\bf 28}'_{+1}+{\bf 224}'_{+1} &
\\[.2ex]
{\bf 21}_{-1}+{\bf 28}_{-1}+\!{\bf 224}_{-1} &
\\[.2ex]
{\bf 7}'_{-3}+{\bf 140}'_{-3} &
\\[.2ex]
{\bf 35}_{-5} &
\\[.2ex]
{\bf 1}_{-7} & 
\end{array}
\label{912G7}
\eea
in which we clearly identify the seven-form flux and the four-form flux 
as the upper two lines.
Also the third line allows for a straightforward interpretation:
the ${\bf 7}+{\bf 140}$ of ${\rm GL}(7)$ corresponds to a tensor with
index structure ${\cal T}^k_{mn}$ and thus precisely to the torsion 
or geometric flux introduced in (\ref{geoflux}).
The resulting four-dimensional theories can thus be obtained
by evaluating the general Lagrangian sketched in section~\ref{subsec:Lagrangian}
(and given in detailed form in~\cite{deWit:2007mt}) for the particular 
embedding tensor $\Theta_M{}^\alpha$, that correspond to the upper lines
of (\ref{912G7}).

Gaugings that are triggered by an embedding tensor corresponding to 
the lower entries in (\ref{912G7}) in contrast do not have a clear origin
in the eleven-dimensional theory. Some of these may however find a 
higher-dimensional interpretation in different compactifications
(such as the type IIB theory considered in the next section) or hint to
the existence of certain non-geometric compactifications 
(see e.g.~\cite{Hull:2004in,Hull:2007jy,DallAgata:2007sr}).

It is important to remember that the restriction of the embedding tensor
to the ${\bf 912}$ representation in fact only represented part of the consistency
constraints imposed in the four-dimensional theory. As we have discussed
in section~\ref{subsec:embedding}, it has to be supplemented with the 
quadratic constraint~(\ref{quadraticconstraint}) in order to define a consistent
gauging.
In the present context this constraint translates into certain bilinear conditions
on the flux parameters $a$, $c_{klmn}$, ${\cal T}^k_{mn}$, that have to be imposed 
for consistency. Indeed, such conditions typically arise in the explicit study
of flux compactifications.
A straightforward way to obtain these bilinear conditions in our framework would
be the explicit decomposition of~(\ref{quadraticconstraint}) under ${\rm GL}(7)$.
The computation may be drastically simplified by making use of a very
compact way to reformulate the quadratic constraint in the four-dimensional
theory. Namely, one may show that for $\Theta_M{}^\alpha$ restricted to
the ${\bf 912}$ representation of ${\rm E}_{7(7)}$, the quadratic 
constraint~(\ref{quadraticconstraint}) can be written in the equivalent form
\bea
\Theta_M{}^\alpha \,\Theta_N{}^\beta\,\Omega^{MN} &=& 0 \;,
\label{quadratic2}
\eea
with the symplectic matrix 
$\Omega^{MN}$ of (\ref{omega}).\footnote{This equivalence can be proven by 
showing that both expressions live in the same ${\bf 133}+{\bf 8645}$
representation of ${\rm E}_{7(7)}$, see~\cite{deWit:2007mt} for details.}
This form of the constraint immediately shows that the embedding tensor
considered as a matrix $\Theta_M{}^\alpha$ has at most half-maximal rank,
i.e.\ that the gauging involves at most 28 out of the 56 possible vector fields. 
More specifically, it guarantees the mutual locality of electric and 
magnetic charges involved in the gauging.

In order to derive possible bilinear relations between the flux parameters,
it is thus useful to explicitly consider the embedding 
tensor $\Theta_M{}^\alpha$ as a matrix according to the decomposition
(\ref{decompE7G7}), (\ref{E7vectors}) which yields
\bea
{\scriptsize
\begin{tabular}{c|cccccc}
 $\Theta_M{}^\alpha$   & 
  ${{\bf 7}}_{-4}$ & 
    ${\bf 35}'_{-2}$ & 
      ${\bf 48}_0$ & 
    ${\bf 1}_0$ & 
  ${{\bf 35}}_{+2}$ & 
  ${\bf 7}'_{+4}$ 
\\\hline
${\bf 7}'_{-3}$ & 
  ${\bf 1}_{-7}$ &
  ${{\bf 35}}_{-5}$ &
  $({\bf 140}'\!+\!{\bf 7}')_{-3}$ & 
  ${\bf 7}'_{-3}$ & 
  $({{\bf 21}}+{{\bf 224}})_{-1}$ & 
   $({{\bf 28}}'\!+\!{{\bf 21}}')_{+1}$ 
\\
 ${{\bf 21}}_{-1}$ & 
   ${{\bf 35}}_{-5}$ & 
  ${\bf 140}'_{-3}$ & 
  $({{\bf 21}}+{{\bf 28}}+{{\bf 224}})_{-1}$ &
  ${{\bf 21}}_{-1}$ & 
  $({{\bf 21}}'\!+\!{{\bf 224}}')_{+1}$ &
    $({{\bf 140}}+ {{\bf 7}})_{+3}$ 
\\
  ${\bf 21}'_{+1}$ &  
   $({\bf 140}'\!+\!{\bf 7}')_{-3}$ & 
  $({{\bf 21}}+{{\bf 224}})_{-1}$ &
   $({{\bf 21}}'\!+\!{{\bf 28}}'\!+\!{{\bf 224}}')_{+1}$ & 
  ${\bf 21}'_{+1}$ & 
  ${{\bf 140}}_{+3}$ & 
     ${\bf 35}'_{+5}$  
\\
  $ {{\bf 7}}_{+3}$ &
   $({{\bf 28}}+{{\bf 21}})_{-1}$ &
 $({{\bf 21}}'\!+\!{{\bf 224}}')_{+1}$ &
   $({{\bf 140}}+{{\bf 7}})_{+3}$ &
  $ {{\bf 7}}_{+3}$ &
  ${\bf 35}'_{+5}$ &
     ${\bf 1}_{+7}$ 
\end{tabular}
}
\quad \;,
\nonumber\\
\label{thetaG7}
\eea
with all entries built from the blocks of~(\ref{912G7}).
In particular, coinciding representations in the bulk of the table 
correspond to the same flux parameters of~(\ref{912G7})
where they all appear with multiplicity one. 
It remains to evaluate the quadratic constraint (\ref{quadratic2})
for this matrix.

To begin with, let us consider the seven-form flux represented
by the ${\bf 1}_{+7}$ which makes a single 
appearance in (\ref{thetaG7}).
The triangular form of this matrix shows that 
(\ref{quadratic2}) is automatically satisfied, i.e.\ the seven-form flux
defines a consistent one-parameter deformation.
It is amusing to note that this particular theory has 
been constructed even before the first maximal gauged supergravity 
of~\cite{deWit:1982ig} was found, however in a form where the gauging 
is hidden in topologically massive two-forms~\cite{Aurilia:1980xj}.

Next, we may study gaugings induced by the four-form flux $c_{klmn}$
which induces two entries in (\ref{quadratic2}). Again, the triangular form
of the resulting matrix~(\ref{thetaG7}) guarantees (\ref{quadratic2}) without further
constraints on $c_{klmn}$. The first non-trivial constraint is met for
gaugings induced by geometric fluxes ${\cal T}^m_{kl}$.
Inspection of the associated matrix~(\ref{thetaG7}) shows that the condition
 (\ref{quadratic2}) has a non-trivial component if the free indices 
 $\alpha$ and~$\beta$ take values in the ${\bf 7}'_{+4}$ and the ${\bf 35}_{+2}$
 --- while the internal index $M, N$ is contracted over the ${\bf 21}_{-1}$.
 The resulting constraint thus lives in the ${\bf 7}'\otimes{\bf 35}$ by which 
 it is entirely determined to be
\bea
{\cal T}^p_{kl}\,{\cal T}^q_{mp} 
+{\cal T}^p_{lm}\,{\cal T}^q_{kp} 
+{\cal T}^p_{mk}\,{\cal T}^q_{lp} 
&=& 0
\;.
\label{qq1}
\eea
One recognizes the standard Jacobi identity, and indeed the ${\cal T}^m_{kl}$
precisely appear as structure constants~(\ref{closure2}) of the local gauge 
algebra~\cite{Kaloper:1999yr,DallAgata:2005ff,DAuria:2005er,Hull:2006tp}.
By similar arguments, one derives the mixed constraint
\bea
{\cal T}^p_{kl}\,c_{mnrs}\,\epsilon^{qklmnrs}-
{\cal T}^q_{kl}\,c_{mnrs}\,\epsilon^{pklmnrs} &=& 0\;,
\label{qq2}
\eea
for gaugings that arise from simultaneous presence of four-form and geometric fluxes.
Both equations (\ref{qq1}) and (\ref{qq2}) have non-trivial solutions.
We have thus identified the relevant flux parameters within the 912
components of the general embedding tensor
and derived the full set of quadratic consistency relations among them.
To complete the analysis it remains to evaluate the full Lagrangian 
of~\cite{deWit:2007mt} and in particular the scalar potential~(\ref{potentialN8})
for these particular embedding tensors, which we will not do here,
see~\cite{DAuria:2005dd,DallAgata:2005fm} for some results.

In principle, the very same analysis can be continued for those gaugings 
induced by the lower lying entries of~(\ref{912G7}).
However, the structure of the matrix (\ref{thetaG7}) shows that 
the resulting quadratic constraints will be more and more involved --- and thus
presumably admit less and less solutions.

%%%%%%%%%%%%%%%%%%%%%%%%%%%%%%%%%%%%%%%
\subsection{IIA/IIB fluxes}
%%%%%%%%%%%%%%%%%%%%%%%%%%%%%%%%%%%%%%%

Finally, we will consider flux compactifications of the ten-dimensional
type IIA/IIB theories on a six-torus $T^6$. 
These compactifications have been exhaustively studied in the literature
in particular in the context of ${\cal N}=2$ and ${\cal N}=4$ supergravity, see 
e.g.~\cite{Giddings:2001yu,Kachru:2002he,DAuria:2003jk,Angelantonj:2003up,Grana:2005jc}
and references therein. 
Type IIB flux compactifications in the context of maximal supergravity
that we sketch here, have been studied in~\cite{deWit:2003hq}.
We should stress that although the presence of fluxes necessarily breaks
maximal supersymmetry, we may still obtain a maximally supersymmetric
four-dimensional Lagrangian, in which supersymmetry is broken
spontaneously in the ground state.

In order to identify the ten-dimensional origin of the four-dimensional fields, 
the relevant subgroups 
of~${\rm E}_{7(7)}$ are the products of the torus ${\rm GL}(6)$ with the global
symmetry groups ${\rm GL}(1)$ and ${\rm SL}(2)$, respectively, of the 
ten-dimensional theories.
The corresponding decompositions~(\ref{decomp}) take the form
\bea
{\rm E}_{7(7)} \;\;&\stackrel{{\rm IIA}}{\longrightarrow}&\;\;
{\footnotesize
\begin{array}{c}
{{\bf 1}}_{\!+3}\\
{{\bf 6}'}_{\!+5/2}\\
{{\bf 20}}_{\!+3/2}\\
{{\bf 15}}_{\!+1}\\
{{\bf 6}}_{\!+1/2}\\\hline
{{\bf 35}}_{0}+{{\bf 1}}_{0}+{{\bf 1}}_{0}\\\hline
{{\bf 6}'}_{\!-1/2}\\
{{\bf 15}'}_{\!-1}\\
{{\bf 20}}_{\!-3/2}\\
{{\bf 6}}_{\!-5/2}\\
{{\bf 1}}_{\!-3}
\end{array}}
\quad\;,
\qquad\quad
{\rm E}_{7(7)}  ~\;\;\stackrel{{\rm IIB}}{\longrightarrow}~\;\;
\begin{array}{c}
({{\bf 1}}_{\!+3},{\bf 2})\\[.2ex]
({{\bf 15}'}_{\!+2},{\bf 1})\\[.2ex]
({{\bf 15}}_{\!+1},{\bf 2})\\[.2ex]\hline
({{\bf 35}}_{0},{\bf 1})+({{\bf 1}}_{0},{\bf 3})\\[.2ex]\hline
({{\bf 15}'}_{\!-1},{\bf 2})\\[.2ex]
({{\bf 15}}_{\!-2},{\bf 1})\\[.2ex]
({{\bf 1}}_{\!-3},{\bf 2})
\end{array}
\quad
\;,\qquad
\label{decompE7G6}
\eea
for the type IIA and IIB theory, respectively.
The second number in the IIB components refers to the 
corresponding ${\rm SL}(2)$ representations.
Using~(\ref{actionGL1}), it is straightforward to identify the 
nilpotent symmetries with positive ${\rm GL}(1)$ charge with the
shift symmetries (\ref{shift1}), (\ref{shift2}) inherited from the
higher-dimensional field content. E.g.\ the highest~${{\bf 1}}_{\!+3}$ 
in both decompositions corresponds to the 
shift~(\ref{shift2}) acting on the scalar obtained by dualizing the two-form
that descends from the ten-dimensional two-form
(which is an ${\rm SL}(2)$ doublet in the IIB theory).
Again these decompositions define a triangular gauge~(\ref{triangular})
with respect to the corresponding Borel subalgebras, in which the 
ten-dimensional origin of the fields is manifest.

In order to identify the ten-dimensional flux parameters within the ${\bf 912}$
components of the embedding tensor we will have to decompose the
latter representation under the different ${\rm GL}(6)$ subgroups
of~(\ref{decompE7G6}).
The result is collected in figures 2{\small A}, 2{\small B},
where the 
vertical axis corresponds to the ${\rm GL}(1)\subset {\rm GL}(6)$ grading
related to rescaling of the six-torus. 
Moreover, in these figures we have made the action
of the S- and T-duality groups explicit.
While the dashed diagonal lines denote orbits under the ${\rm SO}(6,6)$
T-duality subgroup of ${\rm E}_{7(7)}$, orbits under
the ${\rm SL}(2)$ S-duality group correspond to horizontal lines
in the IIB picture.
This allows to directly read off the action of the duality groups
on the various flux parameters.

\begin{figure}[tb]
\hspace*{-2mm}
\begin{minipage}{75mm}
\resizebox{70mm}{!}{\includegraphics{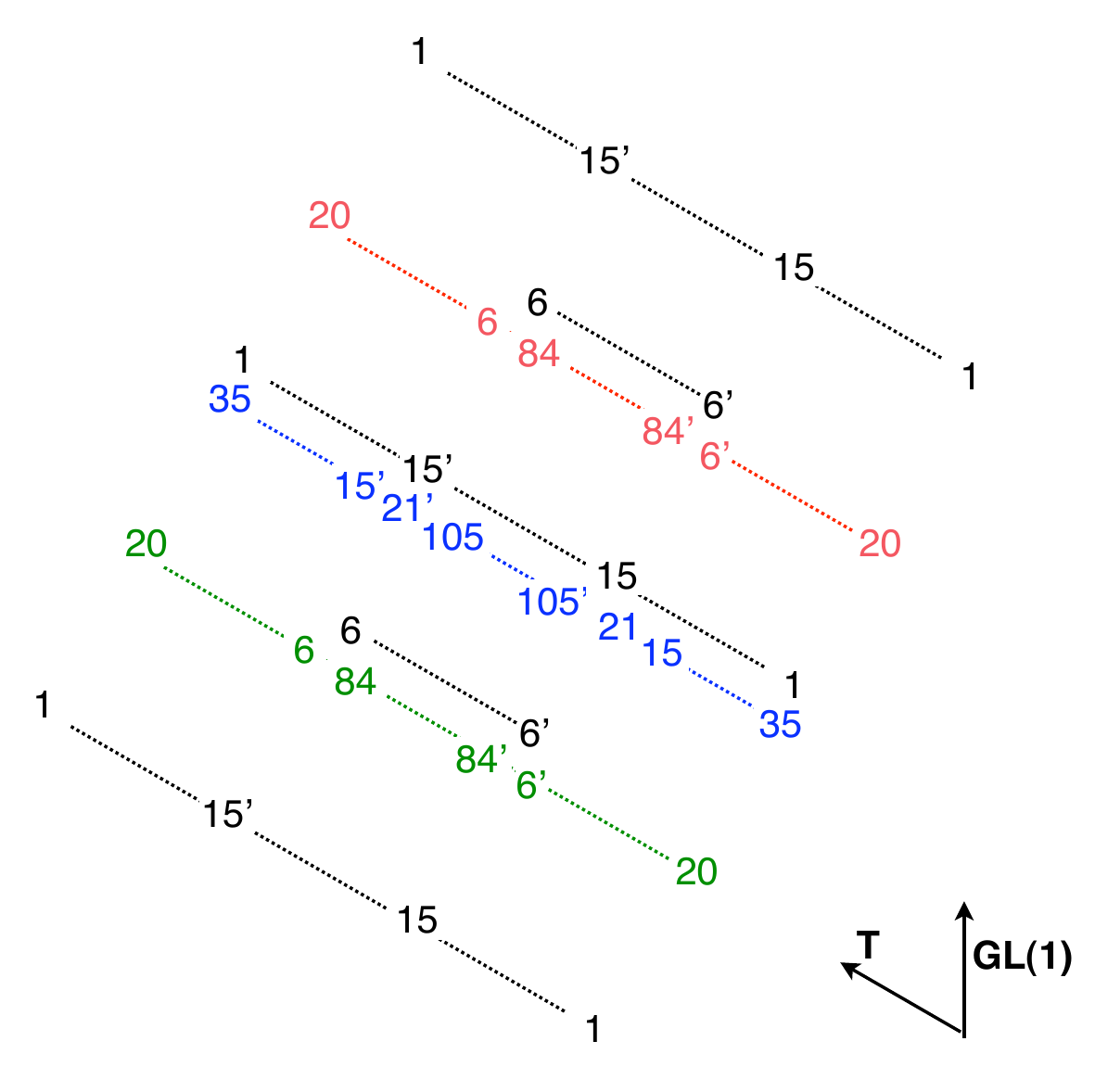}}
\medskip
\addtocounter{figure}{1}

\hspace*{5mm}
\begin{minipage}{60mm}
{{Figure \thefigure {\footnotesize A}:} 
{\small The IIA flux diamond.}}
\end{minipage}
\end{minipage}
\hspace*{8mm}
\begin{minipage}[htbp]{80mm}
\resizebox{73mm}{!}{\includegraphics{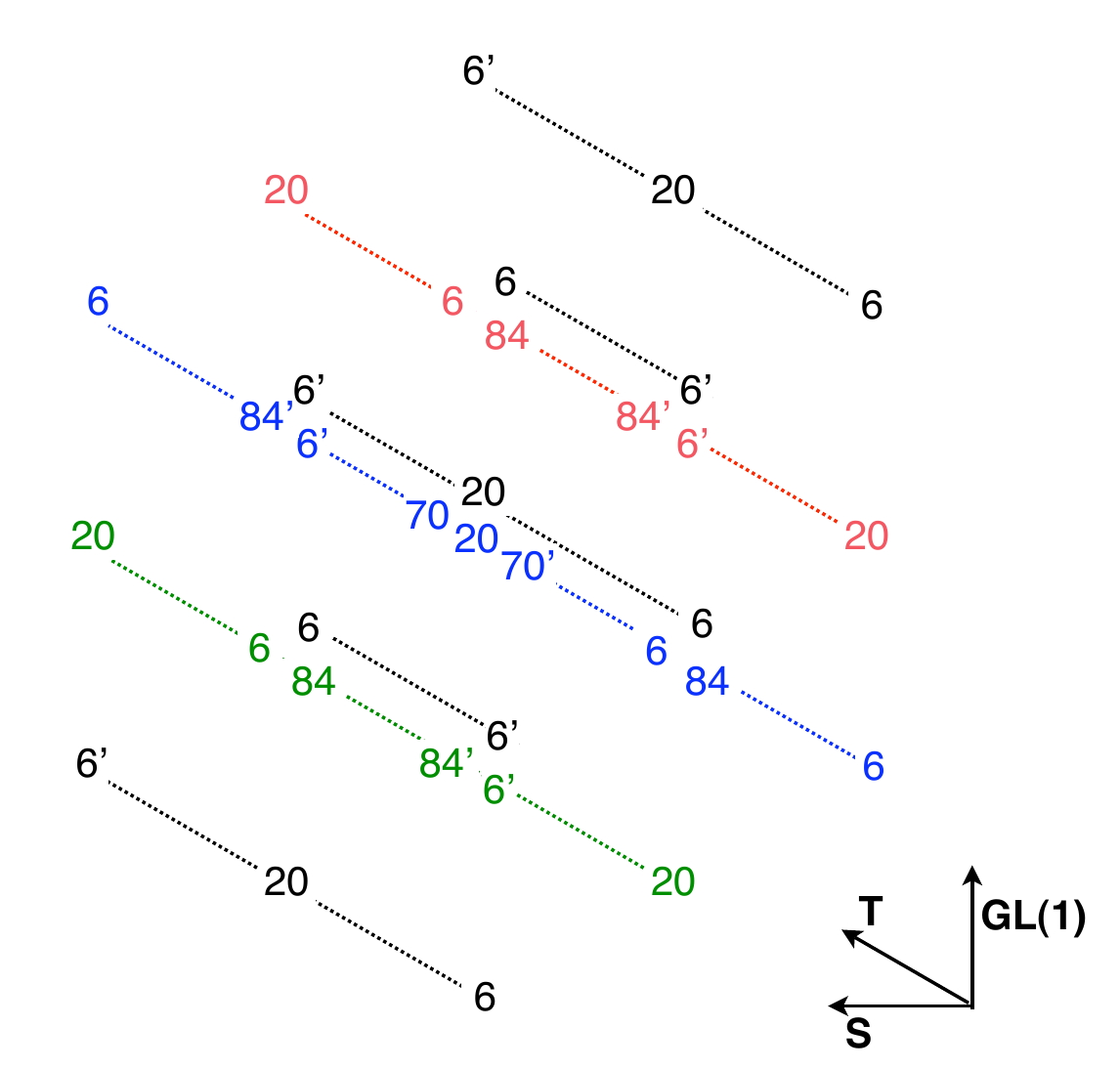}}
\hspace*{8mm}
\begin{minipage}[htbp]{60mm}
{{Figure \thefigure {\footnotesize B}:} 
{\small The IIB flux diamond.} }
\end{minipage}
\end{minipage}
\end{figure}

As in the previous example, the components of the embedding tensor
with highest values of the ${\rm GL}(1)$ grading correspond to $p$-form
fluxes in ten dimensions. In the IIA picture one recognizes the 
six-form, the four-form, the three-form, and the two-form flux, transforming
as ${\bf 1}$, ${\bf 15}'$, ${\bf 20}$ and ${\bf 15}$, respectively,
with their ${\rm GL}(1)$ charges given by~(\ref{actionGL2}).
In the IIB decomposition, the two top rows correspond to the five-form flux ${\bf 6}'$
and the three-form flux doublet $({\bf 20}, {\bf 2})$, respectively.
In both diagrams, the ${\bf 84}+{\bf 6}$ appearing in the following row, denote the parameters
corresponding to geometric flux~${\cal T}^k_{mn}$.
As in the last section, 
evaluating the quadratic constraint~(\ref{quadratic2}) leads to bilinear 
conditions on the flux parameters, such as
\bea
\epsilon^{klmnpq}\,H^\alpha_{klm}\, H^\beta_{npq} 
&=& 0
\;,
\eea
for the three-form flux components in the IIB theory. 
This condition is well known~\cite{Kachru:2002he} and usually modified
by the presence of local sources which explicitly 
break maximal supersymmetry. 

Again, the lower entries in 
figures 2{\small A}, 2{\small B}
correspond to components of the embedding tensor
whose higher-dimensional origin is less obvious.
As can be seen in the figures, they may be reached by subsequent
T- and S-duality transformations starting from known
$p$-form and geometric flux configurations.
An interesting example is the T-duality chain
\bea
H_{kmn} 
~\longrightarrow~
{\cal T}^{k}{}_{mn}
~\longrightarrow~
P_k{}^{mn}
~\longrightarrow~
R^{kmn}
\;,
\eea
corresponding to the diagonal chain 
{\color{Red}
${\bf 20} \rightarrow ({\bf 6}+{\bf 84})
\rightarrow ({\bf 6}'+{\bf 84}')\rightarrow {\bf 20}$},
which has been studied in~\cite{Shelton:2005cf}.
Subsequent application of T-dualities leads from $p$-form
flux $H_{kmn}$ to geometric flux ${\cal T}^{k}{}_{mn}$ and beyond,
to configurations parametrized by tensors
$P_k{}^{mn}$ and $R^{kmn}$ which have been identified as 
so-called non-geometric fluxes.
Furthermore, figure 2{\small B} shows that the parameter $P_k{}^{mn}$
is in fact part of an S-duality doublet $(Q_k{}^{mn}, P_k{}^{mn})$
corresponding to the horizontal pair 
({\color{Blue}${\bf 6}'+{\bf 84}'$}\,, {\color{Red}${\bf 6}'+{\bf 84}'$}).
Indeed, this has been identified and studied in detail in~\cite{Aldazabal:2006up}.
While the usual approach to these non-geometric configurations is
an explicit evaluation of the relevant T- and S-duality transformations,
we see that the covariant scheme discussed in these lectures provides a
framework to construct all four-dimensional theories corresponding to 
the various entries in the flux diamonds of figures 2 in a closed
and manifestly ${\rm E}_{7(7)}$ covariant form.
In particular, all bilinear conditions among the various flux
parameters combine into the single equation~(\ref{quadratic2})
and the full scalar potential for generic (geometric and non-geometric)
fluxes is given by the universal expression~(\ref{potentialN8}).
Upon further orientifold projections, it is then possible to obtain a 
variety of non-maximal theories.
It remains to study the properties of these theories and 
in particular the scalar potential
for the various flux compactifications.
What we have tried to illustrate here is that the covariant formulation
of gauged supergravities provides a universal framework in
which the effective theories associated with particular flux compactifications
can be conveniently constructed and analyzed.

\bigskip
\bigskip

\noindent
{\bf Acknowledgements:}
I wish to thank the organizers and participants
of the RTN Winter School on Strings, Supergravity 
and Gauge Theories at CERN for the opportunity 
to present these lectures and for many comments and suggestions.
It is a great pleasure to thank
Eric Bergshoeff, Gianguido Dall'Agata, Bernard de Wit, Olaf Hohm, 
Axel Kleinschmidt, Arnaud Le Diffon,
Hermann Nicolai, Nicholas Prezas, Diederik Roest, Ergin Sezgin, 
Mario Trigiante, and Martin Weidner,
for the numerous exciting discussions and collaboration
on the topics presented.
This work is in part supported by the Agence Nationale de la Recherche (ANR).

\bigskip
\bigskip
\bigskip
\bigskip

{\small
%\bibliographystyle{Jopt2}
%\bibliography{refs}

\providecommand{\href}[2]{#2}\begingroup\raggedright\endgroup

}

\end{document}